\documentclass[aps,prx,twocolumn,floatfix,superscriptaddress]{revtex4-2}

\usepackage[T1]{fontenc}
\usepackage[separate-uncertainty=true]{siunitx}
\usepackage[mathlines]{lineno}
\usepackage{gensymb}
\usepackage{graphicx}
\graphicspath{{Figures/}}
\usepackage{amssymb}
\usepackage{amsmath}
\usepackage{float}
\usepackage[colorlinks=true, linkcolor=blue,  citecolor=blue,
urlcolor=blue]{hyperref}

\usepackage{multirow}
\usepackage{tabularx}
\usepackage{booktabs}
\usepackage{colortbl}
\usepackage{makecell}
\usepackage{braket}
\usepackage{mathtools}
\usepackage{amsfonts}

\usepackage{nccmath}
    {\end{pmatrix}\end{medsize}}%

\begin{document}

\title[Microsecond-lived quantum states in a carbon-based circuit driven by cavity photons]{Microsecond-lived quantum states in a carbon-based circuit driven by cavity photons}

\author{B.~Neukelmance}
\thanks{These two authors contributed equally to this work.}
\affiliation{Laboratoire de Physique de l’École normale supérieure, ENS, Université PSL, CNRS, Sorbonne Université, Université Paris Cité, Paris, France}
\affiliation{C12 Quantum Electronics, Paris, France}

\author{B.~Hue}
\thanks{These two authors contributed equally to this work.}
\affiliation{Laboratoire de Physique de l’École normale supérieure, ENS, Université PSL, CNRS, Sorbonne Université, Université Paris Cité, Paris, France}
\affiliation{C12 Quantum Electronics, Paris, France}

\author{Q.~Schaeverbeke}
\affiliation{C12 Quantum Electronics, Paris, France}

\author{L.~Jarjat}
\affiliation{Laboratoire de Physique de l’École normale supérieure, ENS, Université PSL, CNRS, Sorbonne Université, Université Paris Cité, Paris, France}

\author{A.~Théry}
\affiliation{Laboratoire de Physique de l’École normale supérieure, ENS, Université PSL, CNRS, Sorbonne Université, Université Paris Cité, Paris, France}

\author{J.~Craquelin}
\affiliation{Laboratoire de Physique de l’École normale supérieure, ENS, Université PSL, CNRS, Sorbonne Université, Université Paris Cité, Paris, France}

\author{W.~Legrand}
\affiliation{Laboratoire de Physique de l’École normale supérieure, ENS, Université PSL, CNRS, Sorbonne Université, Université Paris Cité, Paris, France}
\affiliation{C12 Quantum Electronics, Paris, France}

\author{T.~Cubaynes}
\affiliation{Laboratoire de Physique de l’École normale supérieure, ENS, Université PSL, CNRS, Sorbonne Université, Université Paris Cité, Paris, France}

\author{G.~Abulizi}
\affiliation{C12 Quantum Electronics, Paris, France}

\author{J.~Becdelievre}
\affiliation{C12 Quantum Electronics, Paris, France}

\author{M.~El~Abbassi}
\affiliation{C12 Quantum Electronics, Paris, France}

\author{A.~Larrouy}
\affiliation{C12 Quantum Electronics, Paris, France}

\author{K.F.~Ourak}
\affiliation{C12 Quantum Electronics, Paris, France}

\author{D.~Stefani}
\affiliation{C12 Quantum Electronics, Paris, France}

\author{J.A.~Sulpizio}
\affiliation{C12 Quantum Electronics, Paris, France}

\author{A.~Cottet}
\affiliation{Laboratoire de Physique de l’École normale supérieure, ENS, Université PSL, CNRS, Sorbonne Université, Université Paris Cité, Paris, France}
\affiliation{Laboratoire de Physique et d'Etude des Matériaux, ESPCI Paris, Université PSL, CNRS, Sorbonne Université, Paris, France}

\author{M.M.~Desjardins}
\affiliation{C12 Quantum Electronics, Paris, France}

\author{T.~Kontos}
\thanks{These two authors co-supervised the project.}
\affiliation{Laboratoire de Physique de l’École normale supérieure, ENS, Université PSL, CNRS, Sorbonne Université, Université Paris Cité, Paris, France}
\affiliation{Laboratoire de Physique et d'Etude des Matériaux, ESPCI Paris, Université PSL, CNRS, Sorbonne Université, Paris, France}

\author{M.R.~Delbecq}
\thanks{These two authors co-supervised the project.}
\affiliation{Laboratoire de Physique de l’École normale supérieure, ENS, Université PSL, CNRS, Sorbonne Université, Université Paris Cité, Paris, France}
\affiliation{Laboratoire de Physique et d'Etude des Matériaux, ESPCI Paris, Université PSL, CNRS, Sorbonne Université, Paris, France}
\affiliation{Institut universitaire de France (IUF)}

\begin{abstract}
Semiconductor quantum dots are an attractive platform for the realisation of quantum processors. To achieve long-range coupling between them, quantum dots have been integrated into microwave cavities. However, it has been shown that their coherence is then reduced compared to their cavity-free implementations. Here, we manipulate the quantum states of a suspended carbon nanotube double quantum dot with ferromagnetic contacts embedded in a microwave cavity. By performing quantum manipulations via the cavity photons, we demonstrate coherence times of the order of \SI{1.3}{\micro\second}, two orders of magnitude larger than those measured so far in any carbon quantum circuit and one order of magnitude larger than silicon-based quantum dots in comparable environment. This holds promise for carbon as a host material for spin qubits in circuit quantum electrodynamics.
\end{abstract}

\maketitle

\section{Introduction}\label{sec1}

Spins isolated in nano-scale devices are a well-established platform for solid-state quantum information processing~\cite{lossQuantumComputationQuantum1998, hansonSpinsFewelectronQuantum2007a}. Although elementary multi-qubit systems have been implemented, scaling spin qubit architectures remains an outstanding challenge despite the progress made in recent years~\cite{veldhorstAddressableQuantumDot2014,yonedaQuantumdotSpinQubit2018, philipsUniversalControlSixqubit2022}. The hardware problem at stake in this endeavor constantly triggers the search for new materials and new architectures~\cite{burkardSemiconductorSpinQubits2023}.

One promising architecture, borrowed from superconducting qubits, combines isolated spins and microwave cavity photons in a circuit quantum electrodynamics (cQED) setup~\cite{viennotCoherentCouplingSingle2015, miCoherentSpinPhoton2018a, samkharadzeStrongSpinphotonCoupling2018,landigCoherentSpinPhoton2018a,cubaynesHighlyCoherentSpin2019, harvey-collardCoherentSpinSpinCoupling2022a,yuStrongCouplingPhoton2023,dijkema2025}. The methods used in this context are very appealing for upscaling spin qubit processors. Recent breakthroughs towards this goal include the use of high-impedance cavities to enhance the spin-photon coupling~\cite{samkharadzeStrongSpinphotonCoupling2018,yuStrongCouplingPhoton2023}, which has led to the first implementation of a two-qubit gate between distant spins~\cite{dijkema2025}. However, the coherence times observed for spins in cavities are essentially two orders of magnitude smaller than their cavity-free implementations, limiting the single- and two-qubit gate fidelities.

Among the many potential solutions, carbon nanotubes have shown great promise~\cite{bulaevSpinorbitInteractionAnomalous2008,lairdQuantumTransportCarbon2015}. In principle, carbon nanotube-based circuits display many attractive features: they can form well-defined and tunable quantum dots~\cite{lairdQuantumTransportCarbon2015}, they can be suspended, reducing impact of stray charges, and they can be grown with pure $^{12}$C, offering a nuclear spin-free environment for electronic spin qubits. However, so far they have only been shown to host quantum states with limited coherence of the order of \SI{10}{\nano\second}~\cite{lairdValleySpinQubit2013,peiHyperfineSpinOrbitCoupling2017}.

In this work, we show that carbon nanotube-based quantum devices finally hold to their promise and can serve as an alternative to existing platforms. Our setup displays several features which compare favorably with the state-of-the-art. First, we demonstrate coherent control of quantum states in a carbon nanotube circuit with a coherence time longer by about 2 orders of magnitude than previously observed~\cite{lairdValleySpinQubit2013,peiHyperfineSpinOrbitCoupling2017}. Second, we are able to perform all the basic operations of our carbon-based circuit at zero external magnetic field and elevated temperature $T \approx \SI{300}{\milli\kelvin}$ compared to the driving frequency $f_d \approx \SI{9}{\giga\hertz}$. Third, the quantum states are solely controlled by cavity photons, which fully exploits the possibilities of cQED. Finally, the coherence time of approximately $\SI{1.3}{\micro\second}$ is the largest reported for quantum dots in cavities for any material.

\begin{figure*}[hbt]
\centering
\includegraphics[width=0.95\linewidth, angle=0]{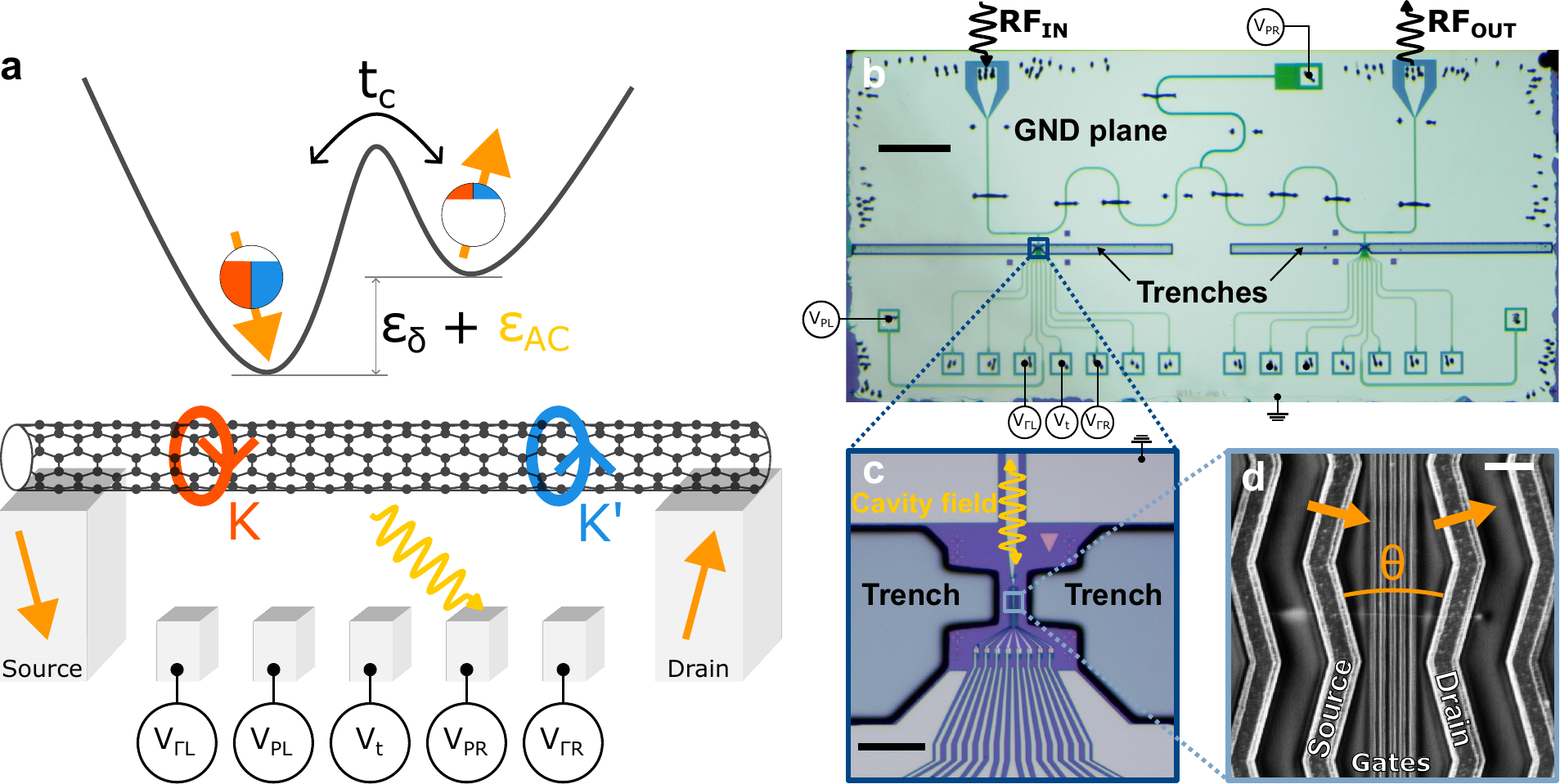}
\caption{\textbf{Schematic and images of the device.} 
\textbf{a.} Schematic of the device. A carbon nanotube is suspended above 5 DC gates and connected to ferromagnetic contacts. The gates allow to form a double well potential in order to trap an electron. $\epsilon_{\delta}$ represents the depth asymmetry between the two wells. The electron can tunnel from one well to the other at a rate $t_c$, thus its wave function is delocalized over the two wells as represented by the filling of each circle. The non-collinear ferromagnetic electrodes impose a different spin quantization axis in each well as shown by the orange arrows. In addition, the electron can be in either valley $K$ (red) or $K'$ (blue). Finally the circuit is coupled to a microwave cavity via the gate $V_{\rm PR}$ which is connected to the central track of the coplanar waveguide resonator, thus inducing a modulation $\epsilon_{\rm AC}$ on the energy detuning between the two wells.
\textbf{b.} Optical image of the sample. On the upper part, the coplanar waveguide resonator is connected to two ports. The IN port is used to both fill the cavity and drive the nanotube circuit. The OUT port is used for reading-out the cavity. In the middle, there are the trenches used to deposit the carbon nanotube. On the bottom part, there are the DC lines used to form the double well potential. The bar is \SI{1}{\milli\metre}.
\textbf{c.} Optical image of a nanotube transfer region showing the large scale view of the electrodes. The scale bar is \SI{50}{\micro\metre}.
\textbf{d.} Scanning electron microscope image of a typical device with a nanotube transferred onto the ferromagnetic electrodes and suspended over the gates. The bar is \SI{1}{\micro\metre} and $ \theta = \pi/6$.
}
\label{SpinQ_architecture}%
\end{figure*}

\section{Results}\label{sec_results}

In order to address long-lived quantum states in a carbon-based circuit via cQED techniques, a spin-photon coupling is engineered~\cite{viennotCoherentCouplingSingle2015}. A suspended carbon nanotube is connected to ferromagnetic electrodes with non-collinear magnetizations and coupled electrostatically to five gate electrodes. One of those is directly connected to the central conductor of a coplanar waveguide (CPW) microwave resonator to induce charge-photon coupling. Additionally, the ferromagnetic electrodes polarize locally the spectrum of the carbon nanotube~\cite{cottetSpinQuantumBit2010a} which induces a synthetic spin-orbit interaction enabling spin-photon coupling~\cite{viennotCoherentCouplingSingle2015}. This is illustrated in the panel a of figure 1. A typical device layout is shown in panels b, c and d. The CPW cavity, visible in panel b, is made out of a \SI{100}{\nano\metre} layer of Nb. Its fundamental resonance frequency at \SI{300}{\milli\kelvin} is $f_c=\SI{6.975}{\giga\hertz}$ with a quality factor $Q=4853$. A close-up on the location of the double quantum dot is shown in panel c. Large trenches are made for the final integration of the carbon nanotube~\cite{cubaynesHighlyCoherentSpin2019, cubaynesNanoassemblyTechniqueCarbon2020a}. The zig-zag shaped ferromagnetic electrodes are designed for creating the non-collinear magnetization and can be seen in panel d on either side of the gates. The ferromagnetic electrodes are made with a Ti(\SI{185}{\nano\metre})/Ni$_{80}$Pd$_{20}$(\SI{35}{\nano\metre})/Pd(\SI{4}{\nano\metre}) stack whereas the gates are made with a Ti(\SI{5}{\nano\metre})/Pd(\SI{50}{\nano\metre}) stack. The gate electrodes having a conducting surface ensure that we can check electrically on the final device that the nanotube is suspended and well isolated from the gates. At the very last step of the nano-fabrication process, a nano-scale transfer of a nanotube grown on a comb via a CH$_4$ based chemical vapor deposition process is performed. It leads to device resistances between \SI{0.5}{\mega\ohm} and \SI{20}{\mega\ohm} at room temperature. The particular device studied here had a room temperature resistance of about \SI{8}{\mega\ohm}.

\begin{figure}[htb]
\centering\includegraphics[width=0.75\linewidth,angle=0]{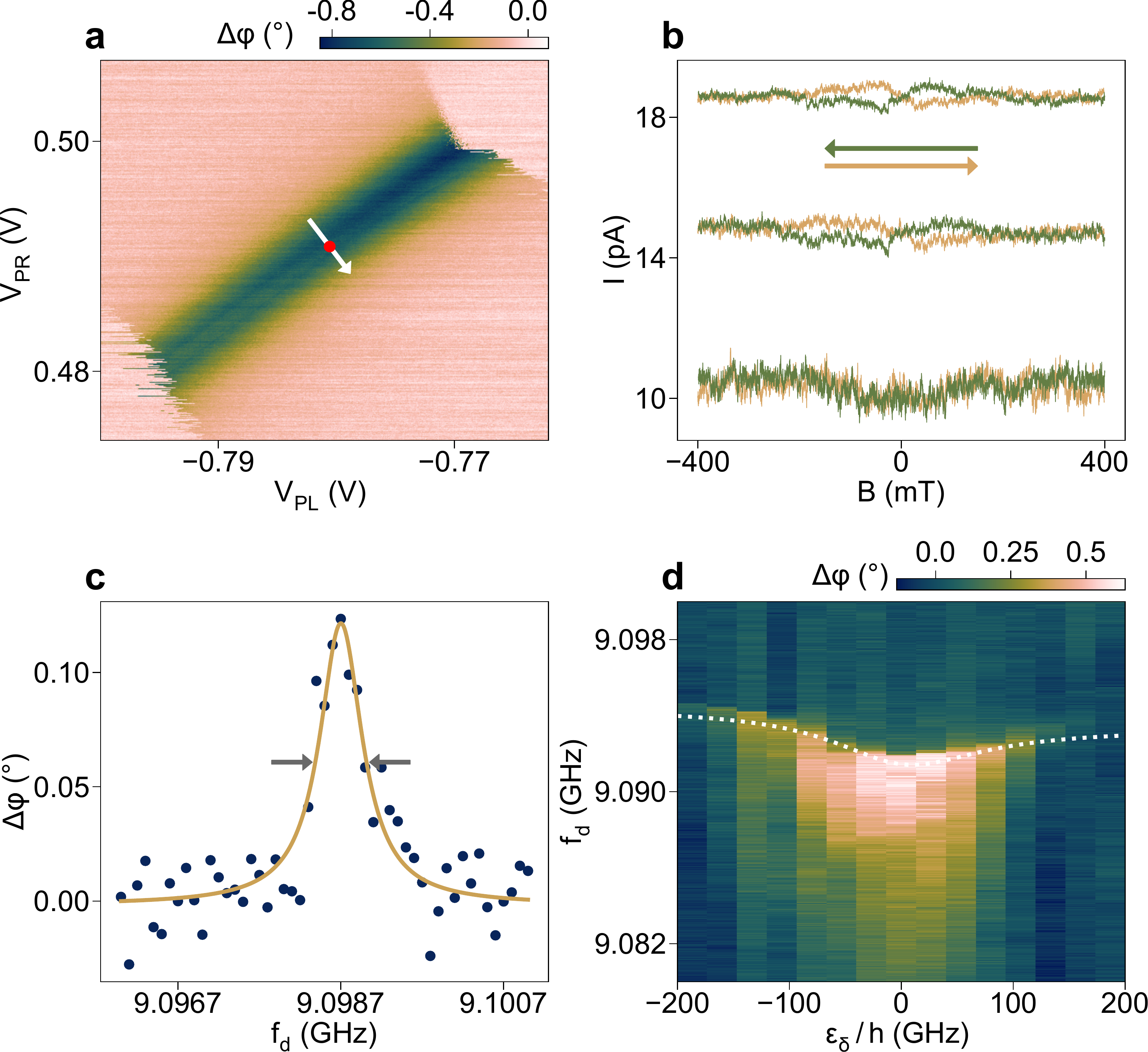}
\caption{\textbf{Spectroscopy and magnetic hysteresis of the circuit.}
\textbf{a.} Continuous wave spectroscopy of the interdot transition studied in the paper. The measured signal is the phase of the microwave signal at $f_c$ as a function of the left and right plunger gate voltages. The arrow indicates the $\epsilon_{\delta}$ axis presented in panel d. The red dot displays the working point $\epsilon_{\delta} = 0$ at which the continuous spectroscopy and the time domain measurements are performed.
\textbf{b.} Transport measurements as a function of the magnetic field showing the characteristic hysteretic magnetoresistance. The arrows indicate which way the magnetic field was swept for each curve. The three pairs of curves correspond to three different plunger gates settings.
\textbf{c.} Two-tone microwave spectroscopy as a function of $f_d$ showing a resonance at $f_r = \SI{9.0987}{\giga\hertz}$. The Lorentzian fit yields a linewidth of \SI{589}{\kilo\hertz}. The grey arrows indicate the full width at half maximum.
\textbf{d.} Dispersion of the resonance frequency with respect to the energy detuning $\epsilon_{\delta}$. The white dashed line is the dispersion predicted by our model. 
}
\label{Qubittransition}
\end{figure}

% \section{Results}\label{sec3}

All the measurements presented in this paper have been made at \SI{300}{\milli\kelvin}. Note that this is an order of magnitude larger than conventional qubit operating temperatures, which has recently been shown to be favorable for mitigating heating from microwave drive~\cite{undsethHotterEasierUnexpected2023}, however we are not in the hot qubit regime~\cite{petitUniversalQuantumLogic2020,yangOperationSiliconQuantum2020}.
The main features of our device can be first characterized via DC transport measurements and continuous wave spectroscopy. The corresponding measurements are shown in figure 2. Panel a shows the phase shift of the microwave tone sent at $f_c$ as a function of the two plunger gate voltages, in the region of interest in this work. We observe the microwave signal characteristic of the transition between two adjacent charge states. Transport measurements could not be performed in the vicinity of this interdot transition--the operating point investigated in this work--because the coupling to the leads was too small. We could however measure current at a different working point in terms of plunger gate voltages. In panel b, the DC current as a function of the magnetic field applied in-planeperpendicularly to the nanotube axis is shown for different working points on the closest bias triangles where significant current is measurable (see Supplementary Discussion~2B). A gate dependent hysteretic signal in the current is observed. It validates spin injection~\cite{Sahoo2005a,viennotCoherentCouplingSingle2015} and therefore the presence of interface exchange fields. Turning to two-tone spectroscopy, we present in panel c the phase contrast probed at $f_c$ as a function of the continuous drive frequency $f_d$. The drive tone is applied via the cavity at an input power $P_{d}=-68$~dBm (all reported microwave powers correspond to the power at the input port of the cavity inside the cryostat, see Supplementary Section~1D for details). A resonance is found at $f_r=\SI{9.0987}{\giga\hertz}$ with a linewidth of \SI{589}{\kilo\hertz} extracted from the Lorentzian fit shown as an orange line in panel c. In this measurement, the cavity frequency shift is due to a change of population of the internal states of the electronic system. This frequency shift in turn yields the measured phase shift. Furthermore, we observe that the resonance frequency varies with respect to the energy detuning $\epsilon_{\delta}$ between the left and right plunger gates, as shown in panel d, confirming that it is a transition of the double quantum dot spectrum. The observed dispersion is strikingly small, a few MHz over hundreds of GHz for $\epsilon_\delta$, yet it is well captured by the theory as shown by the white dashed line~(see Supplementary Discussion~3A). This small dispersion indicates that low charge noise dephasing can be achieved. At the operating point $\epsilon_\delta$ used in the following, the first derivative of the transition frequency with respect to the DQD detuning, $\epsilon_\delta$, vanishes. However, in general, the second derivative does not. A flatter dispersion as the one observed here implies that higher-order derivatives are also minimized, making the system more resilient to charge noise in principle. Overall, these measurements validate that we have indeed implemented the scheme presented in figure~1a.

\begin{figure}[htb]
\centering\includegraphics[width=0.95\linewidth, angle=0]{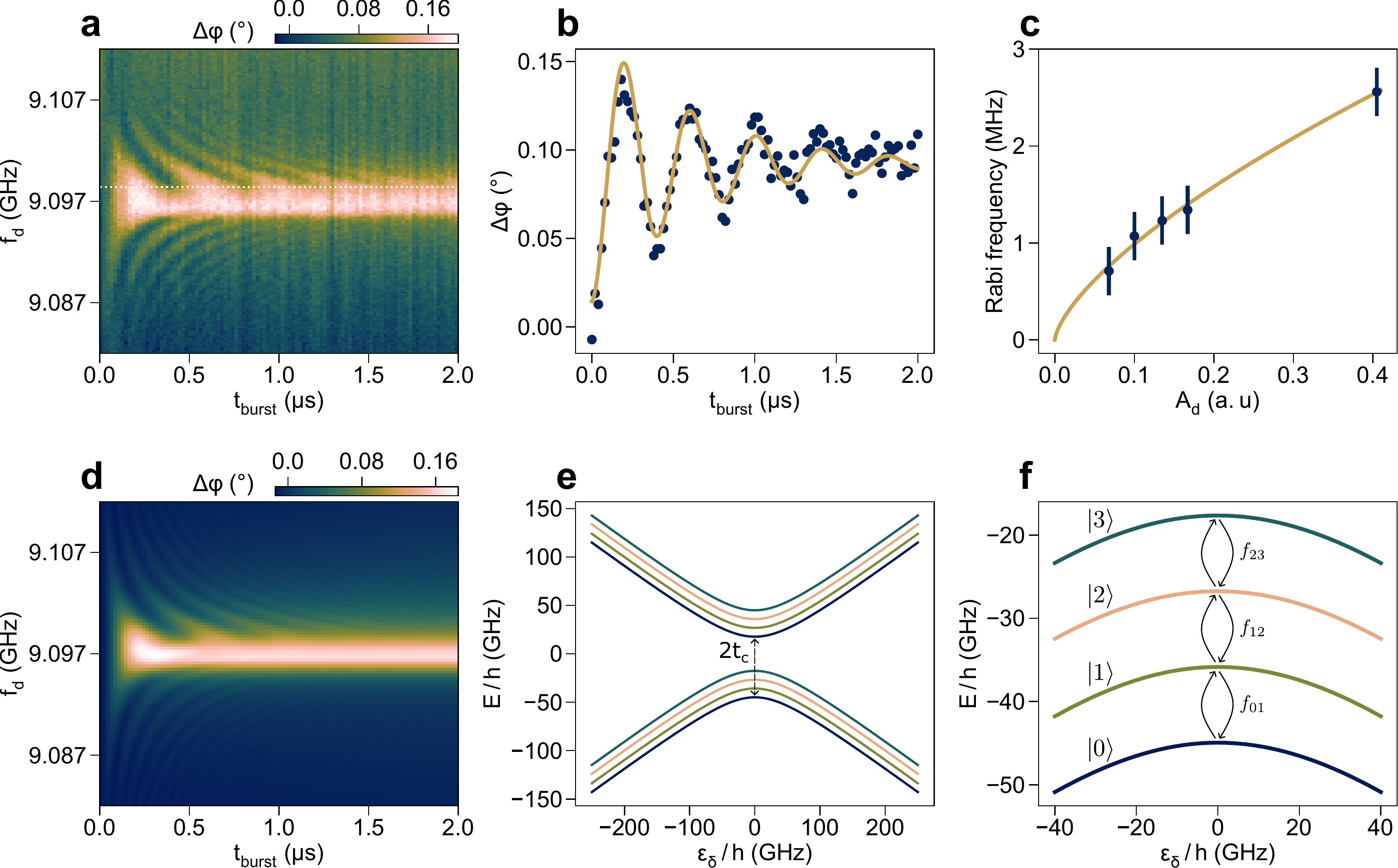}
\caption{\textbf{Rabi oscillations.}
\textbf{a.} Rabi chevrons pattern of the carbon nanotube device. It is observed through the oscillations of the phase contrast $\Delta\varphi$ as a function of the drive frequency $f_d$ and drive time $t_\mathrm{burst}$.
\textbf{b.} Cut in panel a at $f_{d}=\SI{9.0984}{\giga\hertz}$ (white dashed line) showing Rabi oscillations as a function of time fitted by an exponentially decaying cosine.
\textbf{c.} Dependence of the Rabi frequency as a function of the drive amplitude $A_d$ (filled circles). The orange line is a fit to a power law $\Omega_R \propto A_d^{2/3}$. 
\textbf{d.} Simulation of the Rabi chevrons pattern of panel a.
\textbf{e.} Energy spectrum used to model the Rabi chevrons. 
\textbf{f.} Close-up on the first four levels involved in the Rabi oscillations dynamics.
}
\label{Rabichevrons}
\end{figure}

Going to the time domain is crucial to further understand the spectrum and the dynamics of our system. We first perform Rabi oscillations. For this purpose, we excite the system with a frequency $f_d$ close to the resonance found in figure 2c and for a burst time $t_\mathrm{burst}$ up to \SI{2}{\micro\second}, by steps of \SI{20}{\nano\second}. Right after, we read-out the resulting system state using a second pulse at $f_c$~(see Supplementary Discussion~1D). The corresponding measurement is shown in figure 3a. We observe fringes signaling quantum oscillations. These fringes resemble conventional Rabi chevrons observed in many qubit experiments and extend up to almost \SI{2}{\micro\second}. However, there are two main differences here. First, the chevrons for $f_d>\SI{9.0970}{\giga\hertz}$ oscillate slightly slower than those for $f_d<\SI{9.0970}{\giga\hertz}$. Second, there is a horizontal stripe with almost no oscillation contrast very close to the resonance. Nevertheless, the cut at $f_d=\SI{9.0984}{\giga\hertz}$ shown in figure 3b displays well-defined damped oscillations which we fit with an exponentially decaying cosine. We extract a Rabi frequency $\Omega_R/(2\pi)=\SI{2.51}{\mega\hertz}$ and a Rabi decay time $T_\mathrm{Rabi}=\SI{0.59 \pm 0.05}{\micro\second}$.

The Rabi chevrons pattern is very sensitive to the spectrum of the probed system. In particular, our unconventional chevrons pattern tends to indicate a multi-level dynamics~\cite{theryObservationQuantumOscillations2024b}. Furthermore, we observe that the Rabi frequency $\Omega_R$ scales as $A_d^{2/3}$ with $A_d$ the drive amplitude (figure 3c). This scaling strongly suggests a quasi-harmonic spectrum~\cite{claudon2008}. Such a quasi-harmonic ladder can arise from the spectrum of the electronic system. In the simplest picture of a single particle Hamiltonian with two orbital states $\left|L\right\rangle$ and $\left|R\right\rangle$ for left and right dot occupancy, two spin states $\left|\uparrow\right\rangle$ and $\left|\downarrow\right\rangle$, and the two valley states $\left|K\right\rangle$ and $\left|K'\right\rangle$ inherited from the carbon nanotube band structure, the double quantum dot spectrum is made up of 8 levels. From these, one can form two gapped, four-level quasi-harmonic ladders, as illustrated in figure 3e. Phonon modes, naturally present in a suspended CNT, could also lead to an anharmonic ladder via the coupling to the charge degree of freedom of the DQD. In this work, however, we concentrate on a microscopic model based on the electronic spectrum, as it naturally reflects our device design and has already been shown to successfully describe similar devices~\cite{viennotCoherentCouplingSingle2015}. Another reason for focusing on the electronic spectrum is the absence of direct experimental evidence for phonon excitations. We will return to the role of phonons in the context of relaxation in the final section.

We find indeed that with experimentally reasonable parameters for the coupled cavity-DQD system, we can account for both the small dispersion with detuning $\epsilon_\delta$ of the transition, as shown by the white dotted line in figure 2d, and reproduce very well the peculiar Rabi chevrons, as shown in figure~3d. The corresponding spectrum is shown in figures 3e and 3f. It is calculated with the angle $\theta$ between the ferromagnetic contacts set to its lithographically designed value of $\pi/6$, intervalley couplings of the order of \SI{4.5}{\giga\hertz}, exchange fields $E_{L/R}^Z / h$ on the left and the right dots between 9 and 10 \SI{}{\giga\hertz}, a large tunnel coupling $t_c / h \approx \SI{31}{\giga\hertz}$ and a bare charge-photon coupling of $g_0 / (2 \pi) \approx \SI{27}{\mega\hertz}$. All these parameters are estimated by fitting the signal contrast of the interdot transition presented in figure 2a as well as the transition dispersion with detuning presented in figure 2d. The small intervalley couplings indicate that our nanotube is very clean~\cite{lairdQuantumTransportCarbon2015}. Having the intervalley couplings smaller than the exchange fields is essential to obtain the small dispersion in detuning. With these parameters, all eigenstates of the spectrum are fully hybridized between charge, spin and valley at zero detuning. For example, the decompositions of the first two states in the natural basis $\{|LK\uparrow\rangle,\allowbreak|LK\downarrow\rangle,\allowbreak|LK'\uparrow\rangle,\allowbreak|LK'\downarrow\rangle,\allowbreak|RK\uparrow\rangle,\allowbreak|RK\downarrow\rangle,\allowbreak|RK'\uparrow\rangle,\allowbreak|RK'\downarrow\rangle\}$ of our system are $|0\rangle=(-0.376,\allowbreak\, -0.354,\allowbreak\, 0.352,\allowbreak\, 0.331,\allowbreak\, 0.353,\allowbreak\, 0.376,\allowbreak\, -0.332,\allowbreak\, -0.352)$ and $|1\rangle=(-0.352,\allowbreak\, -0.332,\allowbreak\, -0.377,\allowbreak\, -0.354,\allowbreak\, 0.330,\allowbreak\, 0.353,\allowbreak\, 0.354,\allowbreak\, 0.374 )$. Despite this, the transitions between states $|0\rangle$ and $|1\rangle$ and between states $|2\rangle$ and $|3\rangle$ appear to be valley transitions. It is confirmed by an almost flat dispersion of the transition with external magnetic field applied perpendicularly to the nanotube axis both in the experiment and in the simulation~(see Supplementary Discussions~2F and 3D). All other transitions would be spin-valley transitions similar to previously manipulated spin-valley qubits~\cite{peiHyperfineSpinOrbitCoupling2017,lairdValleySpinQubit2013}. In addition, this model and parametrization predicts the coupling of each transition to the cavity mode. By setting a single microwave drive amplitude coefficient, we obtain the Rabi frequencies of each transitions $\Omega_{01}=\SI{2.65}{\mega\hertz}$, $\Omega_{12}=\SI{2.8}{\mega\hertz}$ and $\Omega_{23}=\SI{2.65}{\mega\hertz}$ which then produce the observed Rabi chevron pattern (see Supplementary Discussion~3A). The system is thermally initialized in the thermal equilibrium state with the following calculated populations $(0.77,0.18,0.04,0.01,0,0,0,0)$ in the basis $\{|0\rangle, |1\rangle, |2\rangle, |3\rangle, |4\rangle, |5\rangle, |6\rangle, |7\rangle\}$ (see Supplementary Discussion~3C). Experimentally, we wait for the system to relax in the thermal equilibrium state after each manipulation (see Supplementary Discussion~1D). For the simulation of the Rabi chevron pattern, we used our microscopic model to calculate the relaxation and dephasing rates for all relevant transitions (see Supplementary Discussion~3B). We will return to discussing in more detail decoherence mechanisms in a more model-free approach in the discussion section.

Despite the multilevel dynamics, the observation of well-defined Rabi oscillations at a given drive frequency enables to measure accurately the relaxation time $T_{1}$ and decoherence time $T_{2}^{*}$ of the system. The latter is achieved by Ramsey interferometry, that is, performing two $\pi/2$-pulses separated by a wait time $\tau$ and measuring the resulting state. In our case, a $\pi/2$-pulse is achieved in \SI{100}{\nano\second}. This yields the phase contrast as a function of the drive frequency $f_d$ and the wait time $\tau$ shown in figure 4a. The observed fringes are reproduced in figure 4b using the same model and parameters as the one used for the Rabi chevrons. Moreover, a cut in these fringes at $f_{d}=\SI{9.0984}{\giga\hertz}$ is fitted to a Gaussian decaying cosine which yields $T_{2}^{*}=\SI{1.27 \pm 0.15}{\micro\second}$~(see Supplementary Discussion~2C). This is the highest reported coherence time for a state in a quantum dot in a cavity~\cite{dijkema2025}. Interestingly, we reported in a previous work~\cite{cubaynesHighlyCoherentSpin2019} a similar transition line width of \SI{500}{\kilo\hertz}, inferring a coherence time of the order of \SI{640}{\nano\second}. We therefore see that going to the time domain is essential to access the coherence time of the system, which turns out here to be higher than from the line width estimation. Interestingly, we note that despite the multilevel dynamics of our system, the Ramsey fringes are very similar to the ones of a qubit. This is further confirmed by looking at the Fourier transform (see Supplementary Discussion~2E) where we clearly see that the oscillations frequency depends linearly on the drive frequency detuning. In order to characterize further the dephasing mechanisms in our system, it is instructive to perform a Hahn-echo measurement for which one interleaves a $\pi$-pulse in between the two $\pi/2$-pulses of the Ramsey sequence. This is shown in figure 4c. It allows us to extend the coherence time up to $T_{\rm 2E}=\SI{2.02 \pm 0.21}{\micro\second}$, which can be considered as our $T_2$ time. This value is slightly larger than what has been reported for ensemble of spins on functionalized single wall carbon nanotubes~\cite{chen2023a}. Finally, we get a complete picture of the decoherence by performing a relaxation measurement using a $\pi$-pulse and waiting a time $\tau$ to measure. It yields $T_{1}= \SI{1.12 \pm 0.06}{\micro\second}$ (see figure 4d). The $T_2$ time is thus not far from the $2 \, T_1$ limit which hints at very slow dephasing noise.

\begin{figure}[htb]
\centering
\includegraphics[width=0.95\linewidth,angle=00]{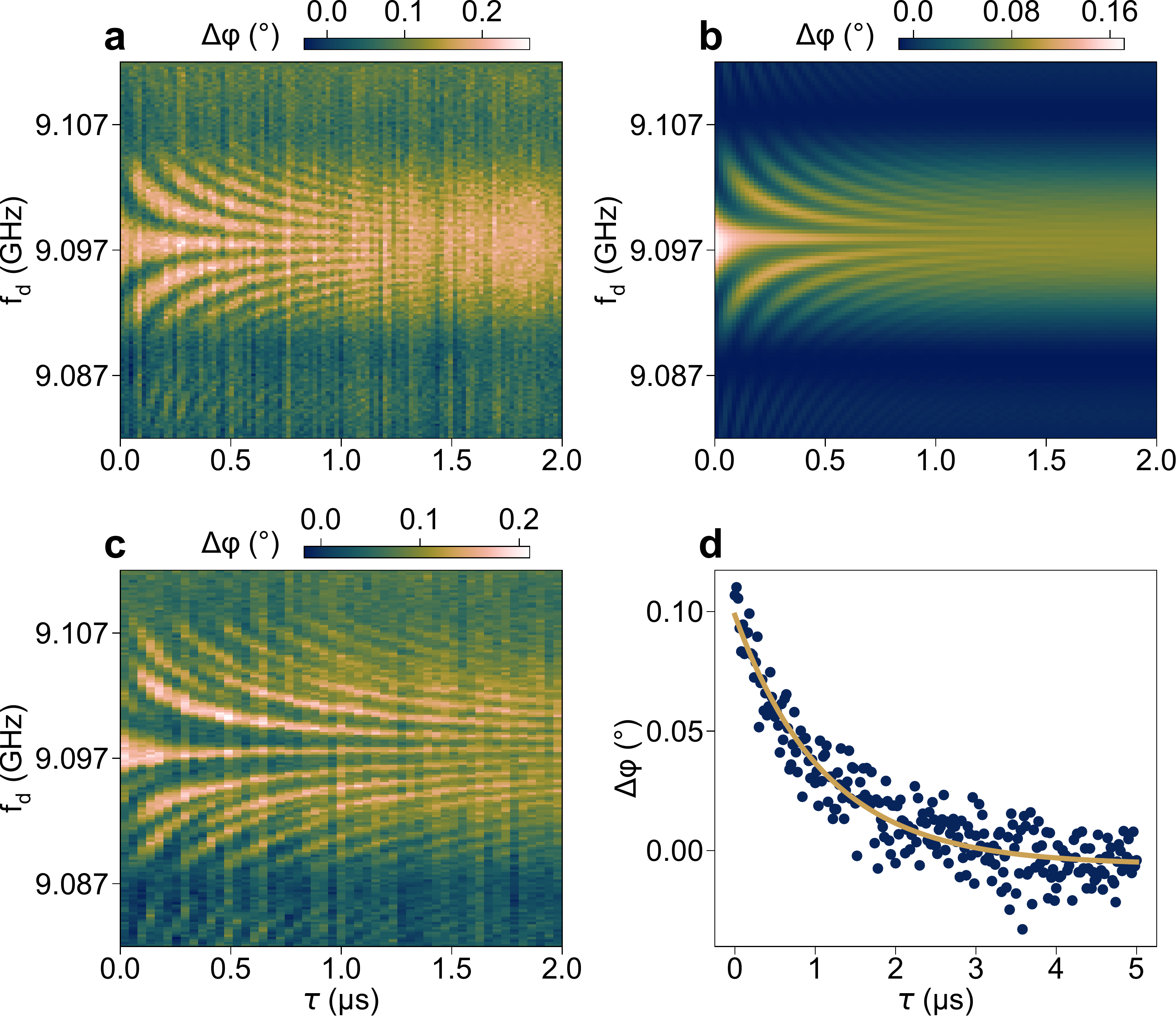}
\caption{\textbf{Decoherence characterization.}
\textbf{a.} Ramsey fringes of the carbon nanotube device. It is observed through the oscillations of the phase contrast $\Delta\varphi$ as a function of the drive frequency $f_d$ and time delay $\tau$ between the two $\pi/2$-pulses of duration $\tau_{\pi/2}=\SI{100}{\nano\second}$.
\textbf{b.} Simulation of the Ramsey fringes pattern of panel a using the level structure displayed in figure~\ref{Rabichevrons}f and using the same parameters as in the figures~\ref{Rabichevrons}d and~\ref{Rabichevrons}e.
\textbf{c.} Hahn-echo fringes.
\textbf{d.} Relaxation measurement yielding $T_{1}= \SI{1.1}{\micro\second}$.
}
\label{Ramsey}
\end{figure}

\section{Discussion}\label{sec4}

Given that we now have all the figures of merit of our system, we can make its decoherence ``budget''. We have shown that a microscopic model (described in Supplementary Discussion~3) can reproduce our observations by incorporating all relevant couplings, relaxation, and dephasing rates of the internal transitions. However, we prefer to remain cautious in drawing strong conclusions regarding the precise values of these estimated parameters as we cannot directly measure experimentally these quantities. Our aim here is therefore to remain as agnostic as possible regarding the details of the model, and to instead focus on data-supported estimates that allow us to discuss each decoherence mechanism in a general manner. Among the processes which can contribute to decoherence, we have radiative processes with the photon bath of the cavity and the phonon bath of the suspended nanotube, charge noise, nuclear spin noise and cotunneling arising from virtual tunneling on and off of the dots levels to the leads. We now detail each process.
\newline
\emph{Purcell relaxation with cavity photons.--} The primary radiative channel with the cavity at $\SI{300}{\milli\kelvin}$ is the Purcell effect since the cavity thermal mean occupation is about 1 photon. The Purcell relaxation rate reads: $\Gamma_{\mathrm{Purcell}} = \kappa g^2 /(\omega_{c}-\omega_{q})^2$, where $\kappa=\omega_c/Q$ is the cavity mode damping rate, $\omega_{c/q}/(2 \pi)$ are the angular frequencies of the cavity mode and the driven transitions respectively, and $g$ is the coupling strength between the transition and the cavity mode. Considering the largest coupling strength in the system, which is the electron-photon coupling $g_0$, we find an upper bound $\Gamma_{\mathrm{Purcell}} / (2 \pi) \leq 2 \times 10^{-4}\SI{}{\mega\hertz}$, which is 4 orders of magnitude below the rates which we measure. Note here that according to our microscopic model, the largest coupling of an internal transition to the cavity mode is \SI{2.5}{\mega\hertz} which would further reduce $\Gamma_{\mathrm{Purcell}}$ by 2 orders of magnitude.
\newline
\emph{Charge noise.--}Experimentally, the transition frequency is sensitive to the detuning $\epsilon_\delta$ as shown in figure 2d. At zero detuning, the first order derivative appears to vanish and one should consider the second order derivative to estimate the charge noise. This is how it is done in the microscopic model (see Supplementary Discussion~3B). However to keep the discussion as model free as possible, we consider the largest dispersion slope observed in figure 2d which is of the order of $4\times 10^{-3} \SI{}{\mega\hertz.\per\micro\electronvolt}$. Taking detuning fluctuations of $\sqrt{\langle \sigma_{\epsilon_\delta}^2 \rangle}\approx \SI{10}{\micro\electronvolt}$, typical for CNT devices in cavities~\cite{viennotOutofequilibriumChargeDynamics2014,bruhatCircuitQEDQuantumdot2018c,cubaynesHighlyCoherentSpin2019}, we obtain an upper bound $\Gamma_{\phi,\mathrm{charge}}/(2\pi) \leq 4\times 10^{-2}\SI{}{\mega\hertz}$ for charge noise dephasing, about one order of magnitude smaller than our observations.
\newline
\emph{Nuclear spin dephasing.--}The presence of $^{13}$C isotope with nuclear spin $I=1/2$ can induce dephasing via the hyperfine coupling to the electron spin. The resulting dephasing is calculated as
$1/\Gamma_{\phi,hf} = T_{\phi,hf} = \hbar/(2 p A) \sqrt{3N /I(I+1)}$, where $p=0.11$ is the natural abundance of $^{13}$C, and $A$ is the hyperfine coupling, theoretically expected to range from $0.1$ to $\SI{1}{\micro\electronvolt}$~\cite{fischer2009}, for which we take a value of $\SI{1}{\micro\electronvolt}$ to be conservative. The number of nuclear spins in the quantum dot is given by~\cite{saito1998} $N=\frac{4 \pi}{3 \sqrt{3} a_{CC}^2}LR$, where $a_{CC}=\SI{1.42}{\angstrom}$ is the carbon-carbon bond length, $L \sim \SI{80}{\nano\metre}$ is the estimated quantum dot length, and $R\sim\SI{1}{\nano\metre}$ is the nanotube radius. This results in an estimate of $\Gamma_{\phi,hf}/(2 \pi) \approx 2\times10^{-2} \SI{}{\mega\hertz}$.
\newline
\emph{Purcell relaxation with phonons.--}As the CNT is suspended, phonons provide a natural relaxation channel to consider. Specifically, we focus on stretching modes, which have a fundamental frequency $\nu_0$ close to the manipulated transitions, given by $\nu_0 L=\SI{12.3}{\giga\hertz.\cdot\micro\metre}$ with $L$ the length of the CNT in \SI{}{\micro\metre}~\cite{mariani2009}. It gives $\nu_0$ between 8.2 and \SI{6.15}{\giga\hertz} for $L$ between 1.5 and \SI{2}{\micro\metre} as designed. Here we cannot resort to using general arguments but have to rely on the microscopic model because the electron-phonon coupling and the phonon modes damping rate are parameters for which there exist no experimental measurement, to our knowledge. We considered Purcell and inverse Purcell effect with up to the fifth harmonics of the phonons modes (see Supplementary Discussion~3B). With a bare electron-phonon coupling of \SI{500}{\mega\hertz} and a damping rate of \SI{2.5}{\giga\hertz} for $L = \SI{1.5}{\micro\metre}$, the model reproduces our measured Rabi chevrons and Ramsey fringes (figures 3d and 4b). However, the simulated global relaxation time $T_1$ (as measured in figure 4d) is $\approx \SI{0.2}{\micro\second}$, which is significantly shorter than the experimentally observed one $\approx \SI{1}{\micro\second}$. This discrepancy indicates that, within the framework of our microscopic approach, phonon-induced relaxation does not appear to be the dominant mechanism. Note that we did not consider other CNT types of phonons like flexural mode, which have fundamental frequencies in the tens of megahertz and can exhibit quality factors in the millions~\cite{moser2014}. Their correspondingly narrow linewidth (on the order of tens of hertz) would be highly unlikely to provide a relevant Purcell relaxation pathway.
\newline
\emph{Cotunneling.--}Finally, we consider cotunneling as a possible source of decoherence. Interestingly, it is known to contribute both to relaxation and decoherence in essentially the same way and amplitude, scaling as $\Gamma_\mathrm{lead}^2/U$ with $\Gamma_\mathrm{lead}$ the coupling to the leads and $U$ the on-site Coulomb interaction~\cite{hartmannDecoherenceChargeStates2002}. We could not perform transport measurement in the phase space region where we perform the manipulation. We therefore could not directly estimate the coupling rate to the leads $\Gamma_{\rm lead}$. Nonetheless, the experimentally observed relationship $T_1 \approx T_2^*$, combined with having ruled out other potential mechanisms, suggests that cotunneling is the dominant source of decoherence in our system.
\newline\newline
In conclusion, our experiment demonstrates the control and readout of microsecond-lived quantum states of a carbon-based circuit embedded in a cQED architecture. This is the longest coherence time reported to date for quantum dots embedded in cavities and is an order of magnitude longer than with equivalent Si-based circuits. The coherence is currently limited by $T_1$. Extending $T_1$ by reducing the coupling to the leads could further increase the coherence in our system to reach the pure dephasing time $T_{\varphi}=\SI{2.9 \pm 0.8}{\micro\second}$ (estimated from the measured $T_1$ and $T_2^*$) which compares favorably with other qubit platform~\cite{Takeda2016}. Remarkably, suspended CNT devices offer a unique platform in which any oxide is far from the quantum circuit and could even be completely oxide-free in future devices. This results in a significant reduction of charge noise due to stray charges, which is known to be one of the main source of decoherence in solid-state qubits. By combining this low charge noise with isotopically purified $^{12}$C nanotubes, carbon circuits could become a promising platform for building spin qubits. Finally, the use of high-impedance resonators to enhance the bare charge-photon coupling should, in future work, enable high-fidelity single- and two-qubit gates far from the Purcell limit.

\vspace*{0.5cm}
\textbf{Acknowledgements:  } WWe acknowledge fruitful discussions with Z.~Leghtas, L.~Bretheau and J.~D.~Pillet. Fundings: This work was supported by the French National Research Agency (ANR) MITIQ (T.K.), the ANR JCJC STOIC (ANR-22-CE30-0009) (M.R.D.), the Emergence project MIGHTY of ville de Paris (M.R.D.), the BPI project QUARBONE (T.K.) and by the ANR through the France 2030 programme through the PEPR MIRACLEQ (ANR-23-PETQ-0003) (M.R.D.).

\textbf{Data availability: }The data that support the findings of this work are available from the corresponding authors upon request.

\textbf{Code Availability: }The codes used for the analysis of this work are available from the corresponding authors upon request.

\textbf{Author contributions: } T.K. and M.R.D. designed and supervised the experiment. B.N. and B.H. performed the measurements. B.N., B.H., G.A. and J.B. fabricated the device with inputs from A.L. for the design. T.C., W.L., M.E., K.F.O., D.S., J.A.S., M.M.D., T.K. and M.R.D. contributed to the development of the CNT nano-assembly technique. L.J. and A.T. developed the acquisition protocols. B.N. B.H., Q.S., A.C., T.K. and M.R.D. analyzed the data and did the theory. B.H., Q.S and M.R.D. did the numerical simulations. B.N., B.H., Q.S., L.J., A.T., J.C., M.M.D., T.K. and M.R.D. contributed to the discussion, interpretation and presentation of the results. B.N. B.H., T.K. and M.R.D. wrote the manuscript with inputs from all the authors.

\textbf{Corresponding authors: } Correspondence should be addressed to
B. Hue (benjamin.hue@phys.ens.fr) and M.R. Delbecq (matthieu.delbecq@phys.ens.fr).

\textbf{Competing interests: } Authors affiliated with C12 Quantum Electronics have financial interest in the company. T.K. and M.R.D. declare equity interest in C12 Quantum Electronics. The remaining authors declare no competing interest.

% \bibliography{Biblio_spin_Ben2.bib}
%apsrev4-2.bst 2019-01-14 (MD) hand-edited version of apsrev4-1.bst
%Control: key (0)
%Control: author (72) initials jnrlst
%Control: editor formatted (1) identically to author
%Control: production of article title (-1) disabled
%Control: page (0) single
%Control: year (1) truncated
%Control: production of eprint (0) enabled
%

%%%%%%%%%% Merge with supplemental materials %%%%%%%%%%
%%%%%%%%%% Prefix a "S" to all equations, figures, tables and reset the counter %%%%%%%%%%

\clearpage
\onecolumngrid
\begin{center}
\textbf{\Large Supplementary materials for "Microsecond-lived quantum states in a carbon-based circuit driven by cavity photons"}
\end{center}

\setcounter{equation}{0}
\setcounter{section}{0}
\setcounter{figure}{0}
\setcounter{table}{0}
\setcounter{page}{1}

\renewcommand{\theequation}{S\arabic{equation}}
\renewcommand{\thesection}{S\arabic{section}}

\renewcommand{\figurename}{Supplementary Figure}
\renewcommand{\tablename}{Supplementary Table}

\title{\Large Supplementary Materials for ``Microsecond-lived quantum states in a carbon-based circuit driven by cavity photons''}

\maketitle

\section{Materials and methods}

\subsection{Wiring}

\begin{figure}[H]
\centering
\includegraphics[width=0.4\textwidth]{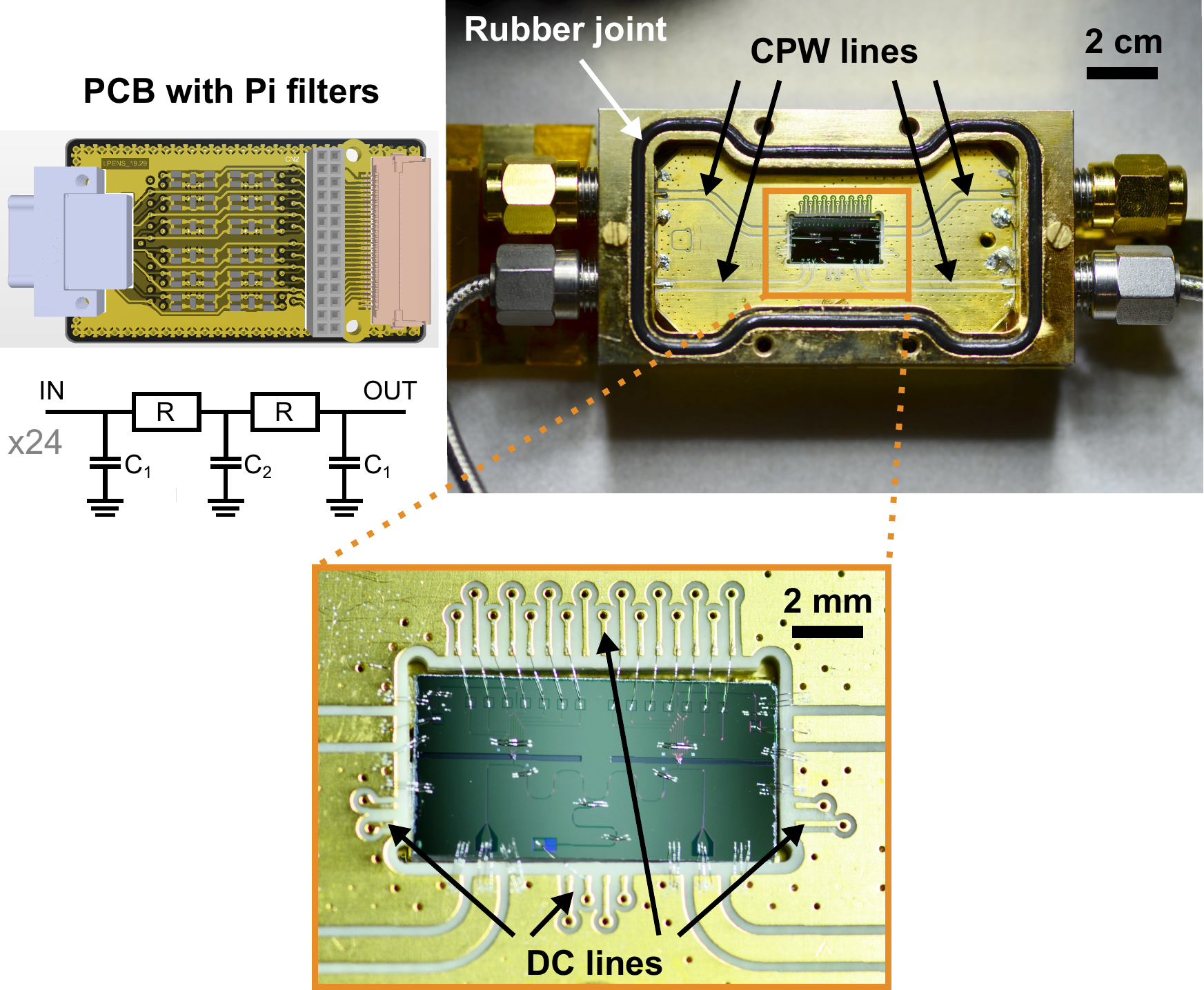}
\caption{\textbf{Circuit chip embedded in the sampler holder.} The PCB contains 24 DC lines and 4 CPW lines for RF signals. Each DC line has a Pi filter: $R = \SI{1}{\kilo\ohm}$, $C1 = \SI{1}{\nano\farad}$ and $C2 = \SI{2.2}{\nano\farad}$. A rubber joint ensures the vacuum tightness when closing the sample holder with its cover inside the stapler.}
\label{sample_holder}
\end{figure}

\begin{figure}[H]
\centering
\includegraphics[width=0.4\textwidth]{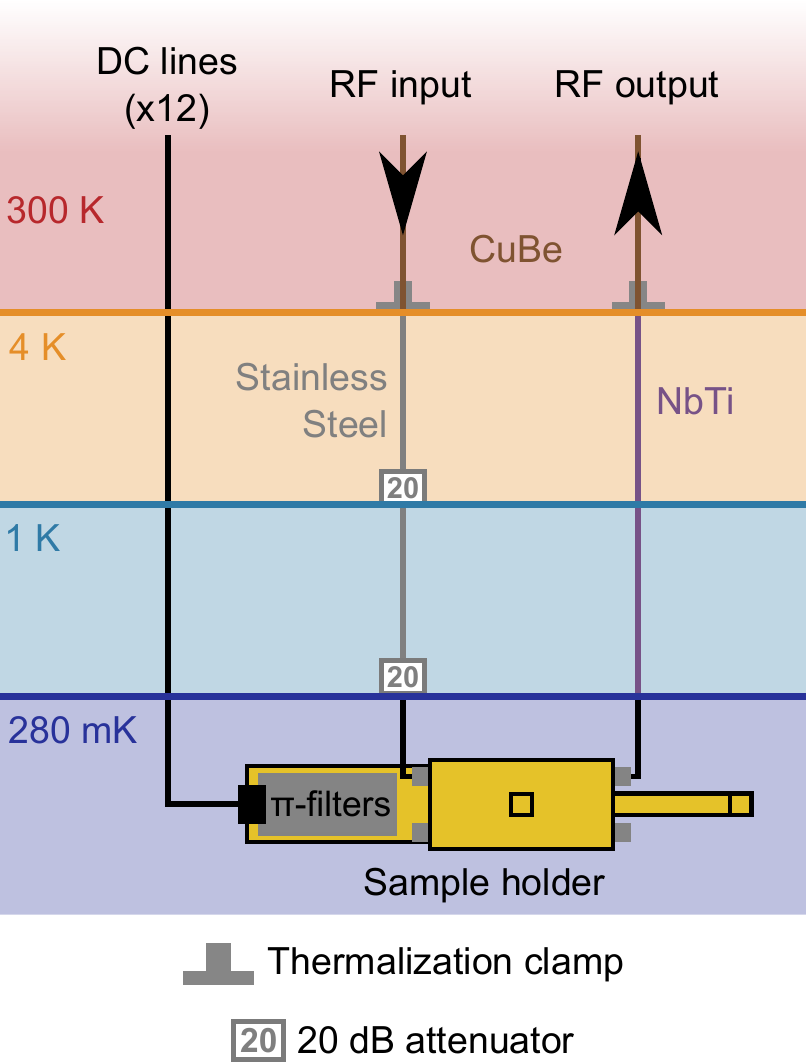}
\caption{\textbf{${}^{3}$He cryostat wiring.} The attenuators thermalize both the outer and inner part of the coaxial cables and reduce thermal noise. The NbTi line provides very few attenuation to the RF output line.}
\label{cryostat_wiring}
\end{figure}

\subsection{DC measurements}

\begin{figure}[H]
\centering
\includegraphics[width=0.7\textwidth]{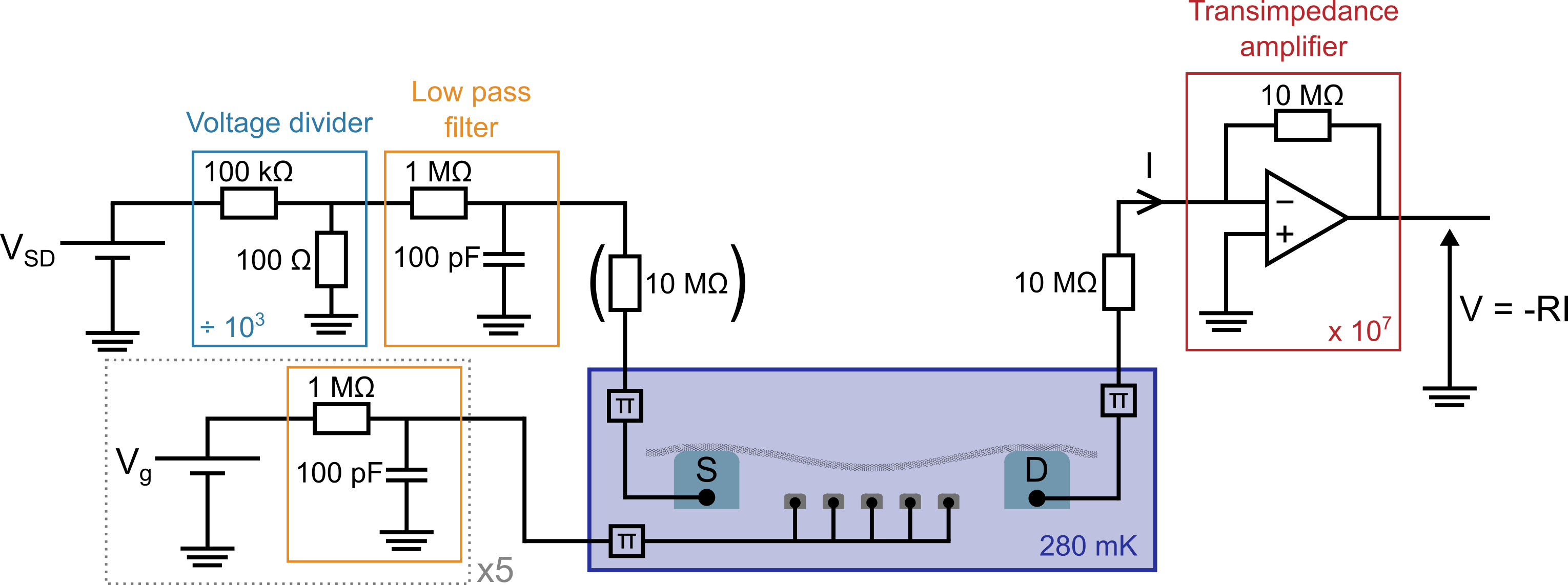}
\caption{\textbf{DC measurements set-up.} The voltage divider reduces the bias voltage $V_{SD}$ by a factor of $10^3$. The current going out of the cryostat is amplified by a factor $10^7$ by a low noise I-V converter.}
\label{DC_setup}
\end{figure}

\subsection{RF measurements}

\begin{figure}[H]
\centering
\includegraphics[width=0.6\textwidth]{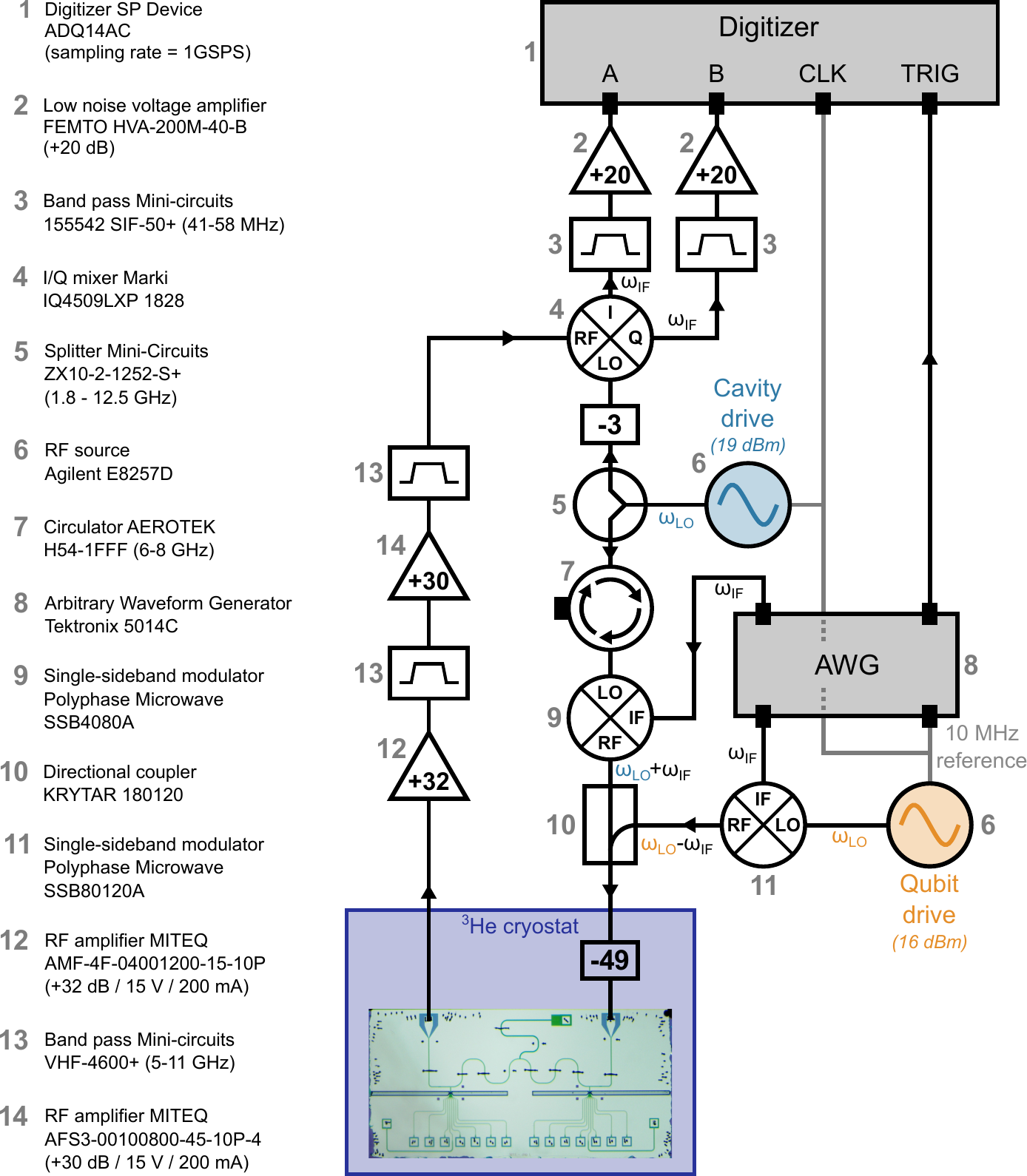}
\caption{\textbf{RF measurements set-up.}}
\label{RF_setup}
\end{figure}

\subsection{Pulse sequences for the RF time domain measurements}

\begin{figure}[H]
\centering
\includegraphics[width=0.7\textwidth]{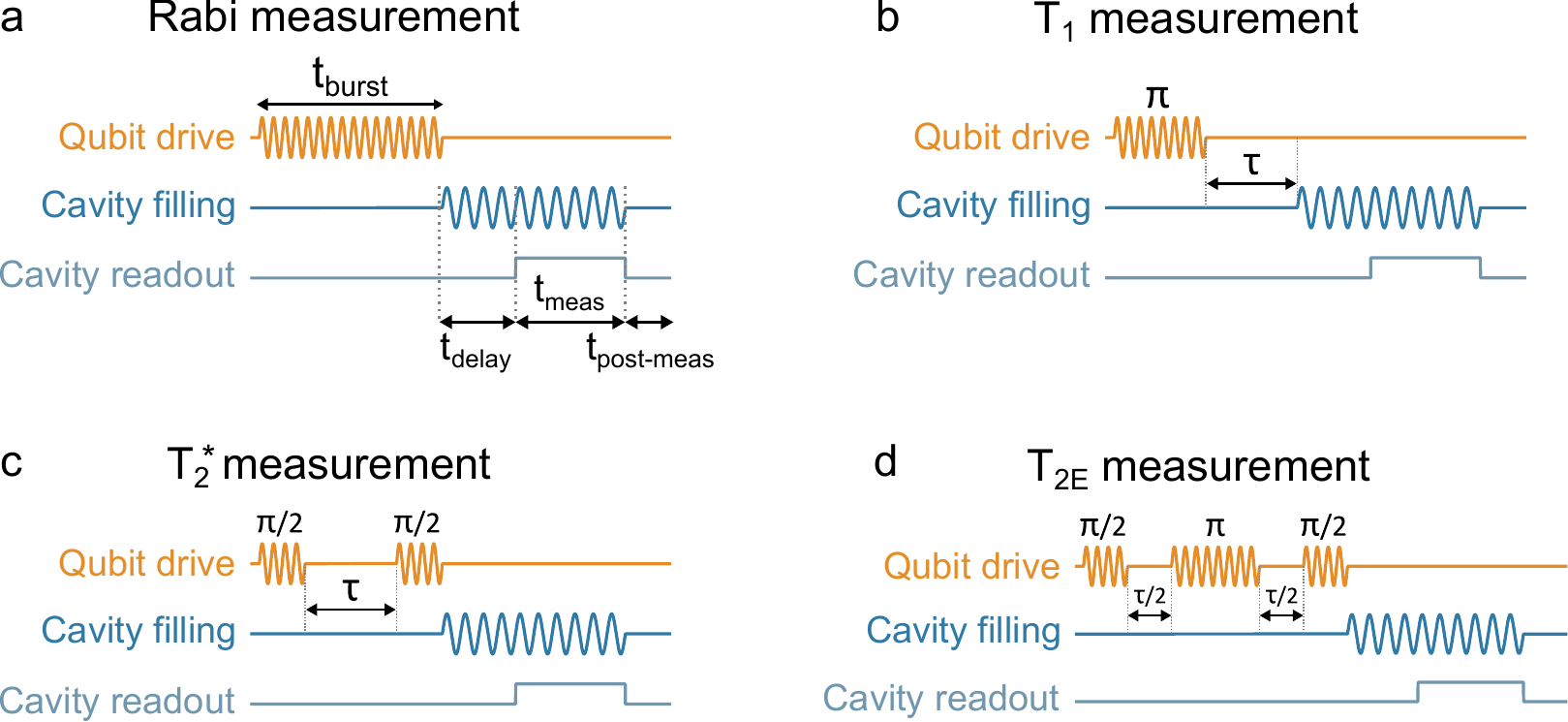}
\caption{\textbf{Pulse sequences.}}
\label{Pulses}
\end{figure}

For each RF time domain measurement presented in the paper, the cavity is measured during $ t_{meas} = \SI{600}{\nano\second}$ after a delay time $ t_{delay} = \SI{250}{\nano\second}$, such that in total the cavity is filled during $ t_{delay} + t_{meas} = \SI{850}{\nano\second}$. We add a post-measuring time $t_{post-meas} = \SI{3}{\micro\second} $ to let both the cavity and the transitions relax in their ground state after the measurement. In order to average the signal acquired, each pulse sequence is repeated $N_{avg} = 150 \, 000$ times. For the Rabi measurements, $t_{burst}$ is varied from $\SI{0}{\micro\second}$ to $\SI{2}{\micro\second}$ by steps of $\SI{20}{\nano\second}$. For the $T_1$ measurement $\tau$ is varied from $\SI{0}{\micro\second}$ to $\SI{5}{\micro\second}$ by steps of $\SI{20}{\nano\second}$. For the $T_2^*$ and  $T_{2E}$ measurements $\tau$ is varied from $\SI{0}{\micro\second}$ to $\SI{2}{\micro\second}$ by steps of $\SI{20}{\nano\second}$. A $\pi$-pulse is achieved in $\SI{200}{\nano\second}$, so logically a $\pi/2$-pulse is achieved in $\SI{100}{\nano\second}$.

Below, we provide a detailed breakdown of the RF power levels used in our measurements. The reported power values correspond to the power at the input port of the cavity inside the cryostat, meaning they already account for the attenuation of the cryostat input line and the entire RF setup.
\begin{itemize}
    \item RF charge stability diagram (Figure~2a of the main text): The cavity was driven with an input power of $-68$~dBm.
    \item Continuous-wave spectroscopy (Figure~2c of the main text): The cavity was driven with an input power of $-75$~dBm for readout, while it was simultaneously driven with an input power of $-68$~dBm for manipulation. The readout power level was deliberately lower to minimize induced broadening (due to the presence of photons in the cavity during the manipulation) and allow for precise characterization of the transition.
    \item Coherent manipulation (Rabi) and coherence measurements (Ramsey, T1, Echo): The cavity was driven with an input power of $-56$~dBm for manipulation readout, while it was afterward driven with an input power of $-58$~dBm for readout. The readout power needed to be larger in this measurement to obtain a sizeable signal, because the cavity was not continuously filled with photons. This readout power was still minimized to reduce measurement induced dephasing.
\end{itemize}

\section{Additional experimental data}

\subsection{Cavity resonance}

\begin{figure}[H]
\centering
\includegraphics[width=0.6\textwidth]{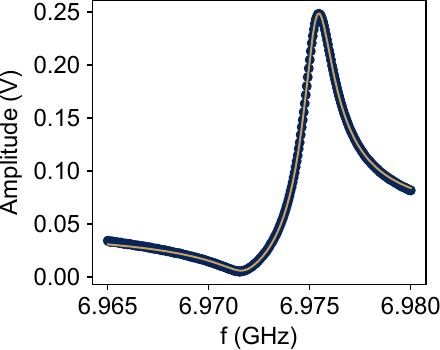}
\caption{\textbf{Cavity resonance.} The filled circles are the data and solid orange line is the fit.}
\label{res_cav}
\end{figure}

The cavity is fitted with $\mid S_{21} \mid$ and $S_{21}$ is defined as follows:

\begin{equation}
S_{21} = \frac{\sqrt{\kappa_{in} \kappa_{out}}}{i 2 \pi (f_c - f) + \kappa / 2 } + i \sqrt{T} e^{i \zeta}
\label{resonance_cav_fit}
\end{equation}

where $f_c$ is resonance frequency of the cavity, $\kappa$ is the linewidth of the cavity, $\kappa_{in}$ and $\kappa_{out}$ are respectively the couplings to the input and output ports of the cavity, $f$ is driving frequency, and $T$ and $\zeta$ are Fano parameters. \\

\begin{table}[ht]
    \centering

    \begin{tabular}{c@{\hskip 1cm}c}
    {Parameters} & {Values}\\
    \arrayrulecolor{black}\Xhline{1pt}

    $f_c$   & \SI{6.975}{\giga\hertz}   \\
    $\kappa$   & \SI{1.437}{\mega\hertz}     \\
    $\sqrt{\kappa_{in} \kappa_{out}}$   & \SI{174.311}{\kilo\hertz}    \\
    $T$     & 0.00226699  \\
    $\zeta$   & 317.369467    \\
    \end{tabular}

    \caption{Fit parameters for the cavity resonance}
    \label{tab:tab_fit_cavity}
\end{table}

\subsection{RF stability diagram}

\begin{figure}[H]
\centering
\includegraphics[width=1\textwidth]{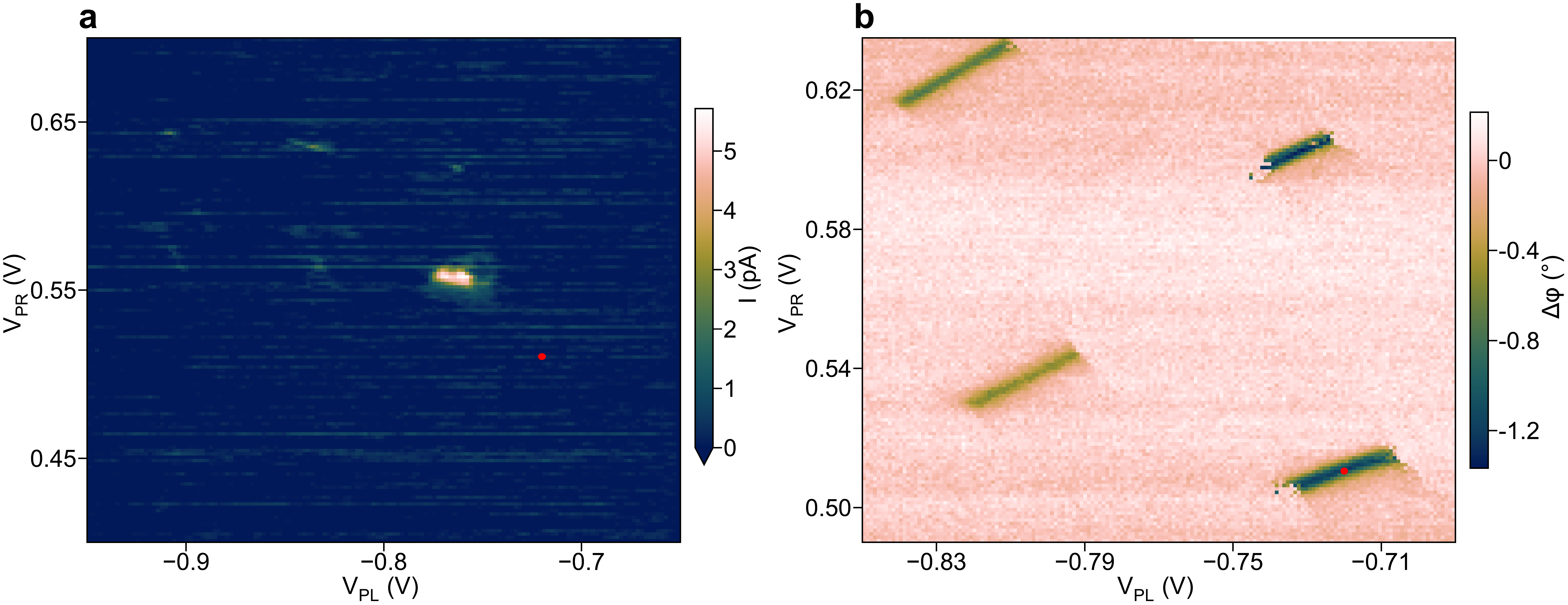}
\caption{\textbf{RF stability diagram.} a) Current measurement of the DQD stability diagram as a function of $V_{\rm PL}$ and $V_{\rm PR}$ at a bias voltage $V_{\rm sd}=\SI{5}{\milli\volt}$. The red dot indicates the position where the transitions have been manipulated. The large current region closeby is the region where the TMR measurements have been performed. b) Charge stability diagram measured through the phase shift $\Delta \phi$ of the cavity field as a function of $V_{\rm PL}$ and $V_{\rm PR}$, showing 4 interdot transitions delimiting a typical charge hexagon. The transitions where manipulated at the location of the red dot, on the lower-right interdot transition.}
\label{RF_stab}
\end{figure}

Supplementary Figure S\ref{RF_stab} presents the RF stability diagram in the area of the interdot transition probed in the manuscript. We see that at the location of the interdot transition where we performed the coherent manipulation presented in the main text (indicated by a red dot in both panels), we could not measure the current. The TMR measurements of Supplementary Figure~S\ref{TMR_wp} where performed at the location of largest current in panel~a.

\subsection{Ramsey oscillations}

\begin{figure}[H]
\centering
\includegraphics[width=0.45\textwidth]{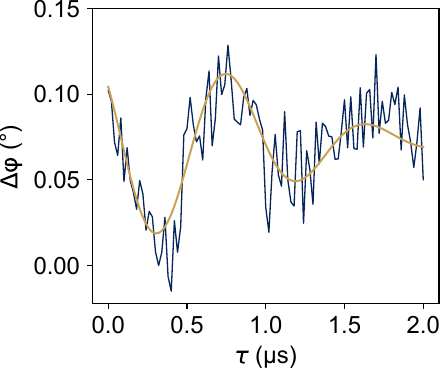}
\caption{\textbf{Ramsey oscillations} }
\label{Ramsey_1D}
\end{figure}

These Ramsey oscillations are obtained by driving at $\SI{9.0986}{\giga\hertz}$. The blue line represents the data and the solid yellow line is a fit with a Gaussian decaying cosine. It yields $T_{2}^{*}=\SI{1.27 \pm 0.15}{\micro\second}$.

\subsection{Hahn-echo oscillations}

\begin{figure}[H]
\centering
\includegraphics[width=0.45\textwidth]{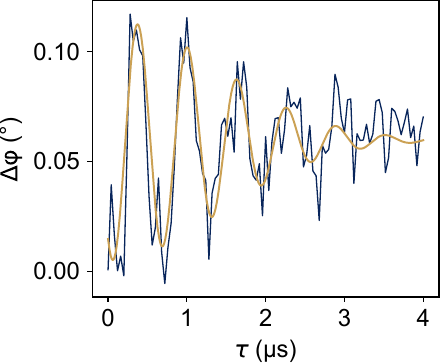}
\caption{\textbf{Hahn-echo oscillations} }
\label{Echo_1D}
\end{figure}

These Hahn-echo oscillations are obtained by driving at $\SI{9.1010}{\giga\hertz}$. The blue line represents the data and the solid yellow line is a fit with a Gaussian decaying cosine. It yields $ T_{\rm 2E} = \SI{2.02 \pm 0.21}{\micro\second} $.

\subsection{Fourier transform of the Ramsey fringes}

\begin{figure}[H]
\centering
\includegraphics[width=0.6\textwidth]{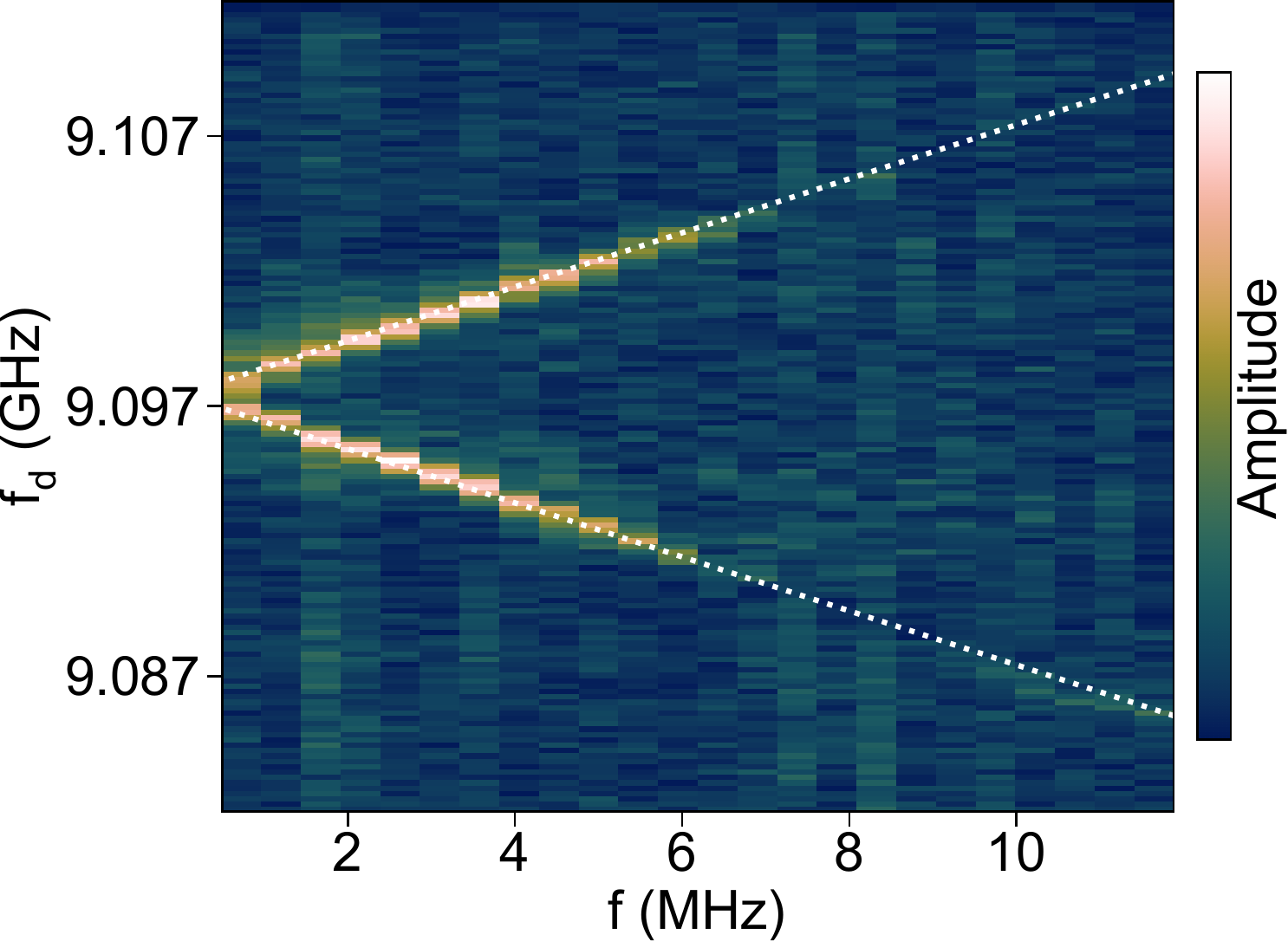}
\caption{\textbf{Fourier transform of the Ramsey fringes.} The white dashed lined plots the following formula : $\mid f - f_0 \mid$ with $ f_0 = \SI{ 9.0974}{\giga\hertz}$. We do observe that the oscillations frequency depends linearly on the drive frequency detuning. }
\label{FFT_ramsey}
\end{figure}

\subsection{Fourier transform of the Hahn-echo fringes}

\begin{figure}[H]
\centering
\includegraphics[width=0.6\textwidth]{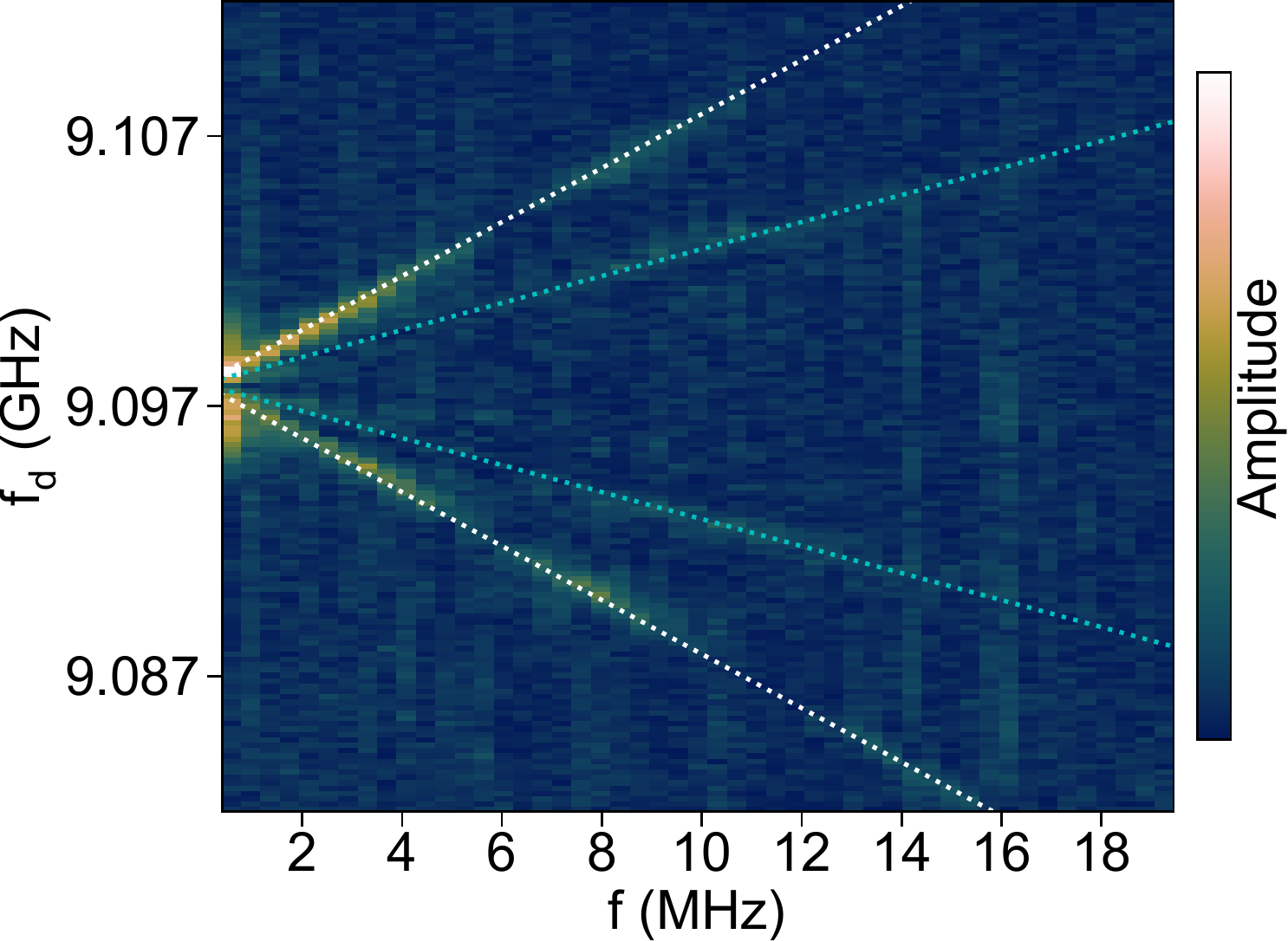}
\caption{\textbf{Fourier transform of the Hahn-echo fringes.} The white dashed lined plots $\mid f - f_0 \mid$ and the blue dashed lined plots $ 2 \times \mid f - f_0 \mid$, both  with $ f_0 = \SI{ 9.0978}{\giga\hertz}$. }
\label{FFT_echo}
\end{figure}

\subsection{Magnetic field dispersion}

Here we present experimental data of the dispersion of the transition measured in two-tone spectroscopy as a function of the external magnetic field $B$. The magnetic field is in-plane and perpendicular to the CNT axis (along the gates defining the DQD). Assuming an electron g-factor $g$ of 2, a change in magnetic field by $\delta B =\SI{10}{\milli\tesla}$ would naively induce a shift in the transition frequency of the order of $\delta f \sim g \mu_B \delta B = \SI{280}{\mega\hertz}$, with $\mu_B$ being the Bohr magneton. This expected shift is two orders of magnitude larger than what we observe experimentally, which leads us to describe the dispersion as (almost) flat. In the case of a magnetic field perpendicular to the CNT, the orbital Zeeman effect is negligible, and we would theoretically expect a flat dispersion.
\newline
However, the actual variation of the transition frequency is more complex due to the full Hamiltonian of the system, which was solved numerically in the \hyperref[sec:Hamiltonian]{Supplementary Discussion~3}. The computed dispersion is presented in Supplementary Figure~\ref{B_simu}, where we observe a frequency variation of approximately \SI{10}{\mega\hertz} over a \SI{100}{\milli\tesla} range and about \SI{1}{\mega\hertz} over \SI{10}{\milli\tesla}, in agreement with our experimental observations. We also note that the simulated dispersion exhibits a small shift of the transition frequency towards lower values with a slight asymmetry, consistent with our measurements.
\newline
Finally, increasing the magnetic field beyond \SI{10}{\milli\tesla} significantly degraded the quality of our measurements. This limitation arises from the deterioration of the superconducting cavity and the already low signal contrast, preventing us from exploring a larger field range.

\begin{figure}[H]
\centering
\includegraphics[width=0.6\textwidth]{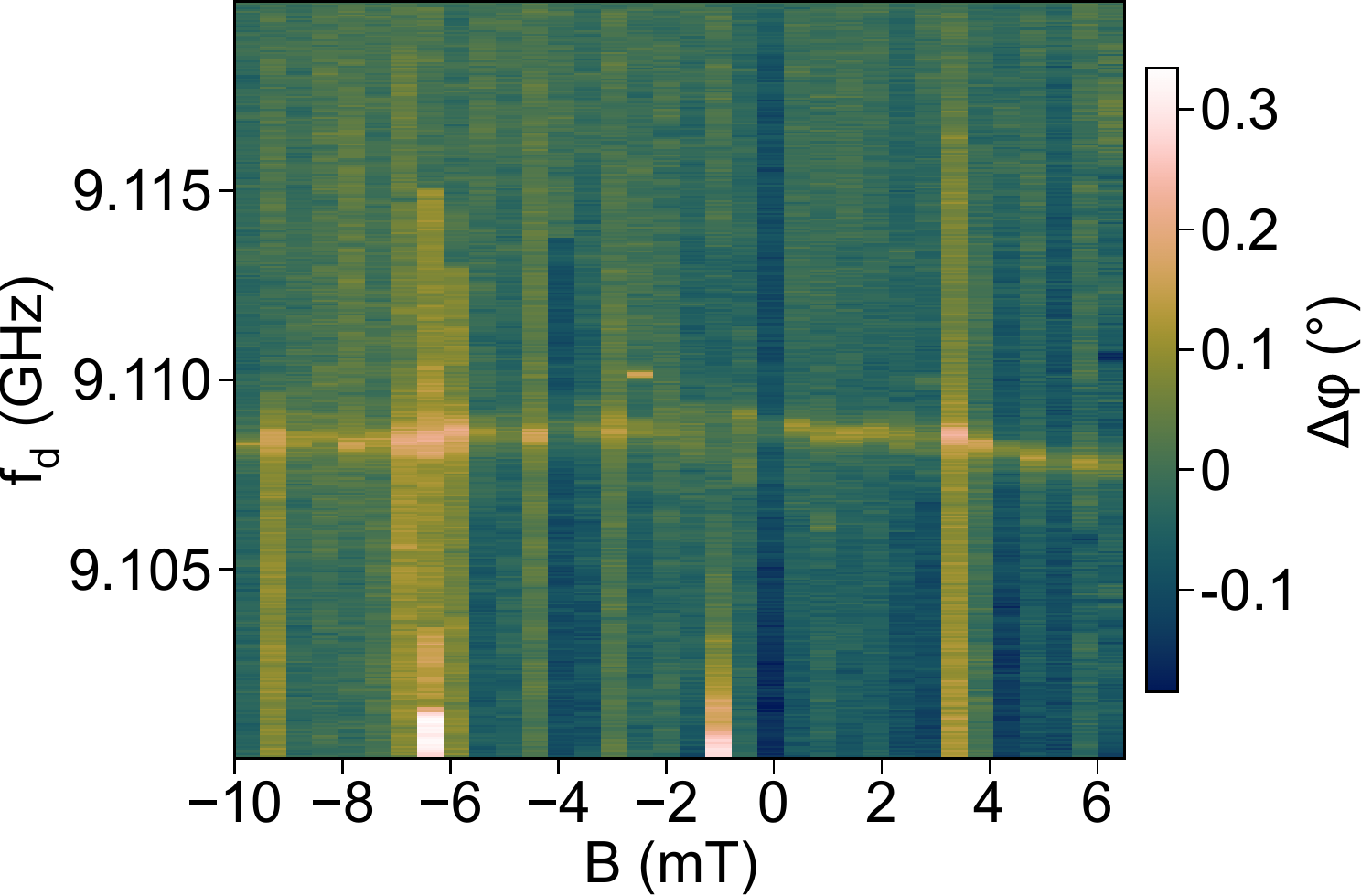}
\caption{\textbf{Magnetic field dispersion.} Dispersion of the transition as a function of magnetic field $B$ which is in-plane and perpendicular to the CNT. The dispersion is almost flat for negative $B$ and shows a slight negative slope at positive $B$. Simulation of this dispersion is presented in Supplementary Fig.~\ref{B_simu}.}
\label{B_field}
\end{figure}

\subsection{Working points for magnetoresistance experiment}

\begin{figure}[H]
\centering
\includegraphics[width=0.6\textwidth]{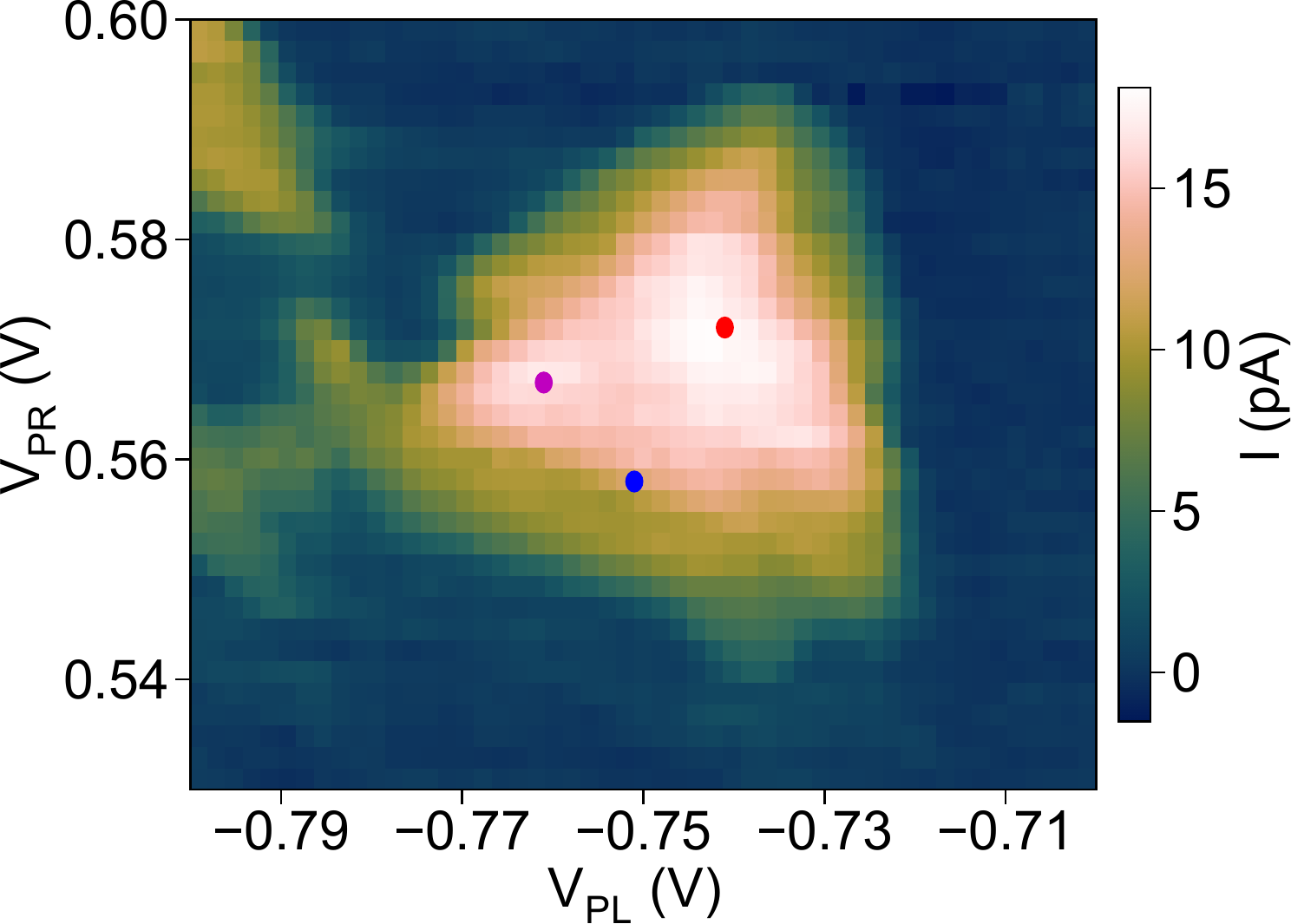}
\caption{\textbf{Working points for magnetoresistance experiment.} The red dot shows the working point for the upper pair of curves in figure 1B, the purple dot shows the working point for the middle pair of curves in figure 1B, and blue dot shows the working point for the lower pair of curves in figure 1B. The bias triangles are measured at $V_{\rm SD}  = \SI{10}{\milli\volt}$. }
\label{TMR_wp}
\end{figure}

\section{Single particle DQD Hamiltonian coupled to a cavity mode}\label{sec:Hamiltonian}

The simulations presented in the main text are based on microscopic modelling of the CNT based double quantum dot circuit coupled to microwave photons~\cite{Viennot2015}.

\subsection{The DQD-cavity Hamiltonian}

The Hamiltonian of the coupled system is

\begin{equation}\label{eq:h_tot}
    \hat{H} = \hat{H}_\mathrm{cav} + \hat{H}_\mathrm{int} + \hat{H}_\mathrm{DQD}.
\end{equation}

The cavity Hamiltonian is

\begin{equation}\label{eq:h_cav}
    \hat{H}_\mathrm{cav} = \hbar \omega_c \hat{a}^\dagger \hat{a}
\end{equation}

with $\omega_c / (2 \pi)$ the cavity frequency and $\hat{a}$ ($\hat{a}^\dagger$) the creation (annihilation) operator of a photon in the cavity mode.

\begin{figure}[htb]
\centering
\includegraphics[width=0.5\textwidth]{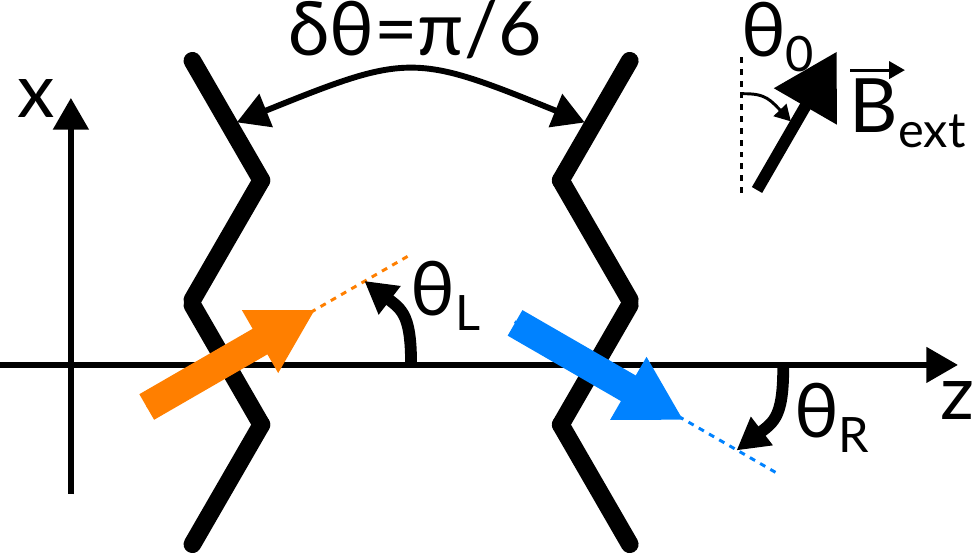}
\caption{Schematic representation of the device showing the the angle parametrizations. The ferromagnetic contacts magnetizations are displayed as orange (left contact) and blue (right contact) arrows making an angle $\theta_L$ and $\theta_R$ with the $z$-axis respectively. The angle between the contacts zig-zag is $\delta \theta = \pi/6$ designed by EB lithography (see figure~1 of the main text). The external magnetic field $\vec{B}_\mathrm{ext}$ has an angle $\theta_0$ with respect to the $x$-axis to account for possible misalignment.}
\label{fig:schematic_magnetization}
\end{figure}

We restrict the Hamiltonian $\hat{H}_\mathrm{DQD}$ of the CNT double quantum dot with non-collinear ferromagnetic contacts to single particle occupancy. We consider the two valleys $K$ and $K^\prime$ in the description. The exchange interaction in the ferromagnetic leads induce effective Zeeman fields $\vec{E}^{\mathrm{Z}}_{\mathrm{L/R,K/K^\prime}}$ which can be different in both dots (due to different contacts) and can differ from $K$ to $K^\prime$. The double quantum dot Hamiltonian reads

\begin{align}\label{eq:h_dqd}
    \begin{split}
    \hat{H}_{\mathrm{DQD}} = & -\frac{1 + \hat{\tau}_z}{2} \frac{\left[\vec{E}^{\mathrm{Z}}_{\mathrm{L,K}} \left( 1 + \hat{\gamma}_z \right) + \vec{E}^{\mathrm{Z}}_{\mathrm{L,K^\prime}} \left( 1 - \hat{\gamma}_z \right) + g_{\mathrm{L}} \mu_\mathrm{B} \vec{B}_\mathrm{ext} \right]}{2} \cdot \vec{\hat{\sigma}} \\
    & -\frac{1 - \hat{\tau}_z}{2} \frac{\left[\vec{E}^{\mathrm{Z}}_{\mathrm{R,K}} \left( 1 + \hat{\gamma}_z \right) + \vec{E}^{\mathrm{Z}}_{\mathrm{R,K^\prime}} \left( 1 - \hat{\gamma}_z \right) + g_{\mathrm{R}} \mu_\mathrm{B} \vec{B}_\mathrm{ext} \right]}{2} \cdot \vec{\hat{\sigma}} \\
    & + \frac{1}{2} \left[ \Delta_\mathrm{L,KK^\prime} \left( 1 + \hat{\gamma}_z \right) + \Delta_\mathrm{R,KK^\prime} \left( 1 - \hat{\gamma}_z \right) \right] \\
    & + \left( t_c \hat{\tau}_x + \epsilon_{\delta} \hat{\tau}_z \right) \hat{\gamma}_0 \hat{\sigma}_0 - \frac{1}{2} g_\mathrm{orb} \mu_\mathrm{B} \left(\vec{B}_\mathrm{ext} \cdot \vec{u}_z \right) \hat{\gamma}_z \hat{\tau}_0 \hat{\sigma}_0,
    \end{split}
\end{align}

with $\vec{E}^{\mathrm{Z}}_{\mathrm{L/R,K/K^\prime}} = E^{\mathrm{Z}}_{\mathrm{L/R,K/K^\prime}} \left[\cos(\theta_\mathrm{L/R}) \vec{u}_z + \sin(\theta_\mathrm{L/R}) \vec{u}_x  \right]$, where $\vec{u}_x,\; \vec{u}_z$ are units vectors along the $x$ and $z$ directions, the $z$ directions being chosen along the CNT axis and $\theta_\mathrm{L/R}$ is the angle between the L/R contact magnetization and the $z$ direction. The parametrization of the angles is described in Fig.~\ref{fig:schematic_magnetization}, giving $\theta_\mathrm{L/R}=\frac{1}{2}\left( \pi - \delta \theta \right)$ where $\delta \theta$ is the angle between the zig-zags of the ferromagnetic contacts, fixed by design to $\delta \theta = \pi / 6$.

In $\hat{H}_\mathrm{DQD}$, $\hat{\sigma}_i, \; \hat{\gamma}_i,\; \hat{\tau}_i$ with $i \in \{0, x, y, z\}$ are the Pauli matrices operators acting in the spin space, the valley space and the L/R space of the double quantum dot respectively, with index $0$ giving the identity matrix in the corresponding subspace. The Landé g-factors $g_{\mathrm{L/R}}$ can in principle be different on both dots, although we fix them equal to 2 in both dots in the following, $\mu_B$ is the Bohr magneton, $\vec{B}_\mathrm{ext}$ is the external magnetic field, $g_\mathrm{orb}$ the orbital g-factor. We account for valley mixing $\Delta_\mathrm{L/R,KK^\prime}$ which can differ in both dots due to spatially varying disorder or curvature of the suspended CNT. Finally the tunnel coupling between the two dots is $t_c$ and the energy detuning between the two levels in each dot is $\epsilon_{\delta}$.

The coupling Hamiltonian between the cavity and the DQD is

\begin{equation}\label{eq:h_int}
    \hat{H}_\mathrm{int} = \hbar g_0 (\hat{a} + \hat{a}^\dagger) \hat{\tau}_z \hat{\gamma}_0 \hat{\sigma}_0,
\end{equation}

where $g_0$ is the bare electron-photon coupling strength.

To reproduce both the dispersion with $\epsilon_{\delta}$ and the time domain measurements, we used two sets of parameters that are close to each other (see table~\ref{table:adjusted_params}). This is motivated by the fact that over the measurement run the frequencies slightly shifted. Then we adjust the spectrum to match the transitions frequencies at zero detuning $\epsilon_{\delta} = 0$ and their dispersion in detuning $\epsilon_{\delta}$ as well as the couplings of each transition to cavity photon to account for the contrast of the Rabi chevrons. The couplings to each transition is $g_{ij}=g_0\Braket{i|\hat{\tau}_z \hat{\gamma}_0 \hat{\sigma}_0 | j}$ where $\Ket{i}$ and $\Ket{j}$ are eigenstates corresponding to eigenvalues of levels $i$ and $j$. With this set of parameters, the couplings matrix $\Braket{i|\hat{\tau}_z \hat{\gamma}_0 \hat{\sigma}_0 | j}$ at zero detuning $\epsilon_{\delta}=0$ is: \\

\begin{equation}\label{eq:g_matrix}
    % \boldsymbol{g} = g_0
        \begin{psmallmatrix}
            0 & 2.4\times 10^{-3} & 8.52\times 10^{-2} & 9.2\times 10^{-3} & 9.9632 \times 10^{-1} & 1.9\times 10^{-3} & 6\times 10^{-5} & 0 \\
            2.4\times 10^{-3} & 0 & 2.5\times 10^{-3} & 8.52\times 10^{-2} & 1.4 \times 10^{-3} & 9.9636 \times 10^{-1} & 0 & 6 \times 10^{-5} \\
            8.52\times 10^{-2} & 2.5\times 10^{-3} & 0 & 2.4\times 10^{-3} & 3 \times 10^{-5} & 0 & 9.9632 \times 10^{-1} & 1.9 \times 10^{-3} \\
            9.2\times 10^{-3} & 8.52\times 10^{-2} & 2.4\times 10^{-3} & 0 & 0 & 3 \times 10^{-5} & 1.4 \times 10^{-3} & 9.9632 \times 10^{-1}\\
            9.9632 \times 10^{-1} & 1.4 \times 10^{-3} & 3 \times 10^{-5} & 0 & 0 & 2.4\times 10^{-3} & 8.52 \times 10^{-2} & 9.2\times 10^{-3} \\
            1.9\times 10^{-3} & 9.9636\times 10^{-1} & 0 & 3\times 10^{-5} & 2.4\times 10^{-3} & 0 & 2.5\times 10^{-3} & 8.52\times 10^{-2} \\
            6\times 10^{-5} & 0 & 9.9636\times 10^{-1} & 1.4\times 10^{-3} & 8.52\times 10^{-2} & 2.5\times 10^{-3} & 0 & 2.4\times 10^{-3} \\
            0 & 6\times 10^{-5} & 1.9\times 10^{-3} & 9.9632\times 10^{-1} & 9.2\times 10^{-3} & 8.52\times 10^{-2} & 2.4\times 10^{-3} & 0
        \end{psmallmatrix}
\end{equation}

\clearpage

With this, we have the Rabi frequencies of the three transitions of the quasi-harmonic ladder given by $\Omega_{01}=\Omega_{23}=\SI{2.65}{\mega\hertz}$ and $\Omega_{12}=\SI{2.8}{\mega\hertz}$, obtained by multiplying the corresponding coupling term by drive amplitude term of value \SI{40.7}{} common to all three transitions. The transition frequencies defined by $f_{ij}=(E_j - E_i)/h$ are $f_{01}=\SI{9098.0}{\mega\hertz}$, $f_{12}=\SI{9096.1}{\mega\hertz}$ and $f_{23}=\SI{9096.0}{\mega\hertz}$.

We can also calculate the decomposition of each eigenstate in the basis $\{\ket{LK\uparrow},\allowbreak\ket{LK\downarrow},\allowbreak\ket{LK'\uparrow},\allowbreak\ket{LK'\downarrow},\allowbreak\ket{RK\uparrow},\allowbreak\ket{RK\downarrow},\allowbreak\ket{RK'\uparrow},\allowbreak\ket{RK'\downarrow}\}$. We have at zero detuning $\epsilon_\delta=0$
\begin{equation}
\begin{pmatrix}
\ket{0} \\
\ket{1} \\
\ket{2} \\
\ket{3} \\
\ket{4} \\
\ket{5} \\
\ket{6} \\
\ket{7}
\end{pmatrix} =
\begin{pmatrix}
-0.376 & -0.354 & 0.352 & 0.331 & 0.353 & 0.376 & -0.332 & -0.352 \\
-0.352 & -0.332 & -0.377 & -0.354 & 0.330 & 0.353 & 0.354 & 0.374 \\
0.362 & -0.322 & -0.383 & 0.344 & -0.323 & 0.360 & 0.344 & -0.385 \\
0.385 & -0.343 & 0.358 & -0.324 & -0.343 & 0.384 & -0.325 & 0.360 \\
0.342 & 0.385 & -0.324 & -0.359 & 0.384 & 0.342 & -0.361 & -0.325 \\
0.321 & 0.362 & 0.344 & 0.382 & 0.36 & 0.322 & 0.384 & 0.344 \\
-0.332 & 0.351 & 0.353 & -0.377 & -0.354 & 0.330 & 0.374 & -0.355 \\
0.353 & -0.376 & 0.331 & -0.353 & 0.376 & -0.354 & 0.351 & -0.332
\end{pmatrix} \begin{pmatrix}
\ket{LK\uparrow} \\
\ket{LK\downarrow} \\
\ket{LK'\uparrow} \\
\ket{LK'\downarrow} \\
\ket{RK\uparrow} \\
\ket{RK\downarrow} \\
\ket{RK'\uparrow} \\
\ket{RK'\downarrow}
\end{pmatrix}
\end{equation}

\begin{table}[ht]
\centering
\scalebox{0.9}{
\begin{tabular}{cc@{\hskip 1cm}cc@{\hskip 1cm}cc}
\multicolumn{2}{c}{Fixed Parameters} & \multicolumn{2}{c}{Adjusted Parameters} & \multicolumn{2}{c}{Adjusted Parameters} \\
\multicolumn{2}{c}{} & \multicolumn{2}{c}{for the dispersion with $\epsilon_{\delta}$} & \multicolumn{2}{c}{for the time domain measurements} \\
\arrayrulecolor{black}\Xhline{1pt}
$\delta \theta$ & $\pi/6$ & $\Delta_\mathrm{L,KK^\prime} / h$ & 4541.2 \SI{}{\mega\hertz} & $\Delta_\mathrm{L,KK^\prime} / h$ & 4543.9 \SI{}{\mega\hertz} \\ [0.8ex]
$\frac{1}{2} g_\mathrm{L} \mu_B$ & 28000 \SI{}{\mega\hertz\per\tesla} & $\Delta_\mathrm{R,KK^\prime} / h$ & 4529.8 \SI{}{\mega\hertz} & $\Delta_\mathrm{R,KK^\prime} / h$ & 4534.7 \SI{}{\mega\hertz}\\ [0.8ex]
% \hline
$\frac{1}{2} g_\mathrm{R} \mu_B$ & 28000 \SI{}{\mega\hertz\per\tesla} & $E^{\mathrm{Z}}_{\mathrm{L,K}} / h$ & 9656.7 \SI{}{\mega\hertz} & $E^{\mathrm{Z}}_{\mathrm{L,K}} / h$ & 9690.1 \SI{}{\mega\hertz} \\ [0.8ex]
 & & $E^{\mathrm{Z}}_{\mathrm{L,K^\prime}} / h$ & 9189.9 \SI{}{\mega\hertz} & $E^{\mathrm{Z}}_{\mathrm{L,K^\prime}} / h$ & 9222.1 \SI{}{\mega\hertz}\\ [0.8ex]
&  & $E^{\mathrm{Z}}_{\mathrm{R,K}} / h$ & 9825,6 \SI{}{\mega\hertz} & $E^{\mathrm{Z}}_{\mathrm{R,K}} / h$ & 9814.6 \SI{}{\mega\hertz} \\ [0.8ex]
 & & $E^{\mathrm{Z}}_{\mathrm{R,K^\prime}} / h$ & 9043,7 \SI{}{\mega\hertz} & $E^{\mathrm{Z}}_{\mathrm{R,K^\prime}} / h$ & 9069.7 \SI{}{\mega\hertz} \\ [0.8ex]
% \hline
 $\omega_c / (2 \pi)$ & 6975 \SI{}{\mega\hertz} & $g_0 / (2 \pi)$ & 27.0 \SI{}{\mega\hertz} & $g_0 / (2 \pi)$ & 27.0 \SI{}{\mega\hertz} \\ [0.8ex]
 $\kappa$ & 1.44 \SI{}{\mega\hertz} & $t_c / h$ & 41020 \SI{}{\mega\hertz} & $t_c / h$ & 31190.0 \SI{}{\mega\hertz} \\ [0.8ex]
\arrayrulecolor{black}\Xhline{1pt}
\end{tabular}
}
\caption{Parameters of the Hamiltonian in \SI{}{\mega\hertz} used for numerical calculations.}
\label{table:adjusted_params}
\end{table}

\subsection{Dephasing and relaxation}

The dephasing is taken into account following the discussion in the supplementary material of ref.~\cite{Cottet2010}. Dephasing for each degree of freedom (charge left/right, spin up/down and valley K/K') is calculated with perturbation theory. For spin and valley, we restrict ourselves to first order while for charge we go to second order to deal with sweet spots where the first order derivative cancels. We have

\begin{align}
\Gamma_{\varphi,ij}^{s}/(2\pi) & = \sqrt{A_s^2 \left|\Braket{j|\hat{\tau}_0 \hat{\gamma}_0 \hat{\sigma}_z | j} - \Braket{i|\hat{\tau}_0 \hat{\gamma}_z \hat{\sigma}_0 | i}\right|^2} \\
\Gamma_{\varphi,ij}^{v}/(2\pi) & = \sqrt{A_v^2 \left|\Braket{j|\hat{\tau}_0 \hat{\gamma}_0 \hat{\sigma}_z | j} - \Braket{i|\hat{\tau}_0 \hat{\gamma}_z \hat{\sigma}_0 | i}\right|^2}
\end{align}

\begin{equation}
\resizebox{.9\hsize}{!}{$
\Gamma_{\varphi,ij}^{c}/(2\pi) = \sqrt{A_c^2 \left|\Braket{j|\hat{\tau}_0 \hat{\gamma}_0 \hat{\sigma}_z | j} - \Braket{i|\hat{\tau}_0 \hat{\gamma}_0 \hat{\sigma}_z | i}\right|^2 + \left[\sum_k \frac{\left|\Braket{k|\hat{\tau}_0 \hat{\gamma}_0 \hat{\sigma}_z | j}\right|^2}{E_k - E_j} - \sum_{k^\prime} \frac{\left|\Braket{k^\prime|\hat{\tau}_0 \hat{\gamma}_0 \hat{\sigma}_z | i}\right|^2}{E_{k^\prime} - E_j} \right]^2 A_c^4}$}
\end{equation}

For combining the dephasing rates associated with each derivative for charge noise we choose the ansatz of computing the square root of the sum of the squares of each contributions. The total dephasing rate for each transition is then the sum of the three contributions $\Gamma_{\varphi, ij} = \Gamma_{\varphi,ij}^{s} + \Gamma_{\varphi,ij}^{v} + \Gamma_{\varphi,ij}^{c}$. The parameters $A_\alpha$ for $\alpha=\{s,\, v,\, c \}$ correspond to the amplitude of fluctuations of the parameters controlling the dephasing. For charge noise it corresponds to the amplitude of fluctuations of the detuning $A_c=\sqrt{\langle \epsilon_\delta \rangle^2}$. For spin dephasing, $A_s$ comprises contributions from the nuclear spins environment and from the fluctuations of the exchange fields. For reproducing the experimental data, we take $A_c=\SI{250}{\mega\hertz}$, $A_s < \SI{10}{\mega\hertz}$ and $A_v = \SI{0.2}{\mega\hertz}$. We note that the value of $A_c$ corresponds to fluctuations of the detuning of standard deviation $\sqrt{\langle \epsilon_\delta \rangle^2} \approx \SI{1}{\micro\electronvolt}$ which is an indication of a very low charge noise, as expected for an ultra-clean carbon nanotube. In the main text however, we keep a more conservative value of $\sqrt{\langle \epsilon_\delta \rangle^2} \approx \SI{10}{\micro\electronvolt}$ to estimate the amplitude of charge noise because we lack independent experimental confirmation of the low detuning fluctuations. The dephasing rates matrix restricted to the relevant subspace $\{\ket{0}, \ket{1}, \ket{2}, \ket{3}\}$ for the dynamics of our system reads in \SI{}{\mega\hertz}

\begin{equation}
    \Gamma_\varphi / (2\pi) =
        \begin{pmatrix}
             & 0.029 & 0.158 & 0.129 \\
            0.029 &  & 0.137 & 0.157 \\
            0.158 & 0.137 &  & 0.030 \\
            0.129 & 0.157 & 0.030 &
        \end{pmatrix}
\end{equation}

For relaxation in the simulations, we consider phonons. At the transition frequency $\sim \SI{9}{\giga\hertz}$, the relevant modes are stretching modes which have a fundamental mode frequency $\nu_0$ given by the relation $\nu_0 L = \SI{12.3}{\giga\hertz\cdot\micro\metre}$ with $L$ the length of the CNT in $\SI{}{\micro\metre}$~\cite{mariani2009}. Due to the nanoassembly technique, we do not know exactly the location of the CNT along the gates. Depending on its location on the zig-zag contacts, the length of the CNT suspended over the gate electrodes varies typically between \SI{1.5}{} and \SI{2}{\micro\metre} giving $\nu_0$ between \SI{8.2}{} and \SI{6.15}{\giga\hertz}. Relaxation is accounted for by a radiative Purcell process. As we work at a temperature comparable to the transitions frequencies, the thermal occupancy $n_{ij}^{\rm th}$ is non vanishing, hence inverse Purcell effect which excite lower states to higher states are non zero and must be taken into account. We have

\begin{align}
    \Gamma_{1,ij}^{\downarrow}/(2\pi) & = \kappa_\nu/(2\pi) \left(\frac{g_{ij}^{\nu}/\sqrt{n}}{\nu_n - f_{ij}} \right)^2 (n_{ij}^{\rm th} + 1) \\
    \Gamma_{1,ij}^{\uparrow}/(2\pi) & = \kappa_\nu/(2\pi) \left(\frac{g_{ij}^{\nu}/\sqrt{n}}{\nu_n - f_{ij}} \right)^2 n_{ij}^{\rm th}
\end{align}

with $\nu_n=n \nu_0$ the frequency of the $n$-th harmonic and $n_{ij}^{\rm th}=\frac{1}{e^{h f_{ij}/k_B T} - 1}$. Here we take the scaling that the charge-phonon coupling $g_{ij}^{\nu}$ scales as $1/\sqrt{n}$~\cite{mariani2009}. The charge-phonon coupling matrix elements are the same as the charge-photon matrix elements given in \eqref{eq:g_matrix}. The bare charge-phonon coupling $g_0^\nu$ is however different than the bare charge-photon coupling. $g_0^\nu$ is expected to be large and is a free parameter in the model. The phonon mode damping rate $\kappa_\nu$ is the third free parameter and as its scaling with the harmonic number is not clearly predicted, we kept it constant. For the simulation we considered up to $n=5$ harmonics so that all transitions of the quasi-harmonic ladder have a contribution. Varying the length $L$ of the CNT, we find that to reproduce the data, we have very similar relaxation matrix with $\Gamma_{1,02}^{\downarrow}$ and $\Gamma_{1,13}^{\downarrow}$ dominating the relaxation process. Restricting again to the $\{\ket{0}, \ket{1}, \ket{2}, \ket{3}\}$ subspace, the relaxation matrix for $L=\SI{1.5}{\micro\metre}$, $g_0^\nu=\SI{500}{\mega\hertz}$ and $\kappa_\nu = \SI{2.5}{\giga\hertz}$ is in \SI{}{\mega\hertz}

\begin{equation}
    \Gamma_1 / (2\pi) =
        \begin{pmatrix}
             & 6 \times 10^{-3} & 8.4 \times 10^{-1} & 3 \times 10^{-3} \\
            1 \times 10^{-3} &  & 7 \times 10^{-3} & 8.4 \times 10^{-1} \\
            4.6 \times 10^{-2} & 2 \times 10^{-3} &  & 6 \times 10^{-3} \\
            4 \times 10^{-5} & 4.6 \times 10^{-2} & 1 \times 10^{-3} &
        \end{pmatrix}
\end{equation}

The parameters used to compute the dephasing and relaxation rates for the simulation of the Rabi chevron and the Ramsey fringes presented in the main text are summarized in table~\ref{table:params_dephas_relax}

\begin{table}[ht]
\centering
\begin{tabular}{cc@{\hskip 2cm}cc}
\multicolumn{2}{l}{Dephasing} & \multicolumn{2}{c}{Relaxation} \\
\arrayrulecolor{black}\Xhline{1pt}
$A_c$ & $\SI{250}{\mega\hertz}$ & $g_0^\nu$ & \SI{500}{\mega\hertz} \\ [0.8ex]
$A_s$ & \SI{10}{\mega\hertz} & $\kappa_\nu$ & \SI{2500}{\mega\hertz} \\ [0.8ex]
% \hline
$A_v$ & \SI{0.2}{\mega\hertz} & $L$ & \SI{1.5}{\micro\metre} \\ [0.8ex]
\arrayrulecolor{black}\Xhline{1pt}
\end{tabular}
\caption{Parameters used to calculate the dephasing and relaxation rates.}
\label{table:params_dephas_relax}
\end{table}

With these parameters, we can reconstruct the decoherence rate matrix in \SI{}{\mega\hertz}

\begin{equation}
    \Gamma_2 / (2\pi) =
        \begin{pmatrix}
             & 0.032 & 0.578 & 0.130\\
            0.029 &  & 0.140 & 0.577\\
            0.181 & 0.138 &  & 0.033\\
            0.129 & 0.180 & 0.030 &
        \end{pmatrix}.
\end{equation}

We can also calculate the corresponding cooperativity of each transition to the cavity field, defined as $C_{ij}=\frac{4 g_{ij}}{\kappa \Gamma_{2,ij}}$, giving

\begin{equation}
    C_{01} = 0.4,\; C_{02} = 26.2,\; C_{03} = 1.4,\; C_{12} = 0.1,\; C_{13} = 26.2,\; C_{23} = 0.4
\end{equation}

We can also estimate a model ``agnostic'' cooperativity from our measurement. The phase contrast of the two-tone spectroscopy or the Rabi chevron is given by $\Delta \phi = 2 g^2 / (\kappa \Delta_{qd})$ with $\Delta \phi = 0.1 \times \pi / \SI{180}{\radian}$ and $\Delta_{qd}$ the qubit drive angular frequency detuning. From this we can estimate a global cooperativity $C=\Delta_{qd} \Delta \phi / \Gamma_2^*\approx 25$.

\subsection{Cavity readout}

The transmission $S_{21}$ of microwave field through the dqd-cavity system is given by~\cite{Cottet2017}:

\begin{equation}\label{eq:readout}
    S_{21} = \frac{\sqrt{\kappa_1 \kappa_2}}{\Delta_\mathrm{cd} - \mathrm{i} \frac{\kappa}{2} - \Xi},
\end{equation}

with $\Delta_\mathrm{cd} = \omega_{\rm c} - \omega_{\rm d}$ the detuning between the cavity and the drive angular frequencies, $\kappa/(2\pi)$ the cavity linewidth. The total susceptiblity $\Xi$ is given by

\begin{equation}
    \Xi = \sum_{ij} \frac{g_{ij}^2 (n_j - n_i)}{\omega_{ij} - \omega_{\rm d} - \mathrm{i} \frac{\Gamma_{ij}}{2}},
\end{equation}

with $n_{i,j}$ the population of level $i, j$, $\omega_{ij}=(E_j - E_i)/\hbar$ are transitions angular frequencies and $\Gamma_{ij}$ is the decoherence rate associated with the transition between states $\ket{i}$ and $\ket{j}$ given by $\Gamma_{ij}=\frac{\Gamma_{1,ij}}{2} + \Gamma_{\varphi, ij}$. The thermal equilibrium populations are given by the Boltzmann weight

\begin{equation}
    n_i = \frac{e^{-E_i / k_BT}}{\sum_k e^{-E_k / k_BT}},
\end{equation}

with $E_i$ the energy of the $i$-th level. In this work we have $T=\SI{300}{\milli\kelvin}$. For simulations of the Rabi and Ramsey experiments, the initial state is the thermal equilibrium state $(0.77,0.18,0.04,0.01,0,0,0,0)$. The calculated populations at each evolution time are then used in equation~\eqref{eq:readout} to calculate the phase of the cavity field. To reproduce the phase contrast observed experimentally, we have to manually change a single parameter from the value computed from the Hamiltonian. This parameter is $g_{03}$ which must be increased by a factor approximately 40. We checked that using this value to calculate the relaxation rates due to phonons reproduces the same simulation given that we slightly reduce $\kappa_\nu$ to \SI{1.5}{\giga\hertz}. This highlights that parameters used to compute the relaxation rates with phonons are not discriminant.

\subsection{Simulated dispersions}

\begin{figure}[H]
\centering
\includegraphics[width=0.45\textwidth]{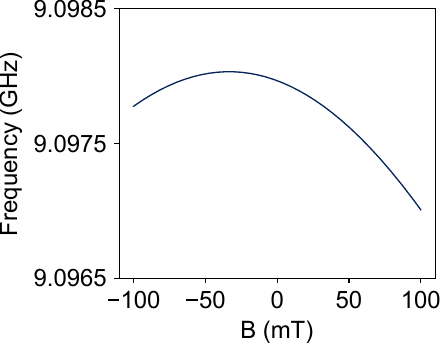}
\caption{\textbf{Simulated dispersion of the 01 transition frequency with respect to the external magnetic field. } It is computed with the parameters used to reproduce the time domain measurement. The very small dispersion in magnetic field is compatible with the experimental observation of Supplementary Fig.~\ref{B_field} and with a transition that is mostly valley-like in this configuration.}
\label{B_simu}
\end{figure}

\begin{figure}[H]
\centering
\includegraphics[width=0.45\textwidth]{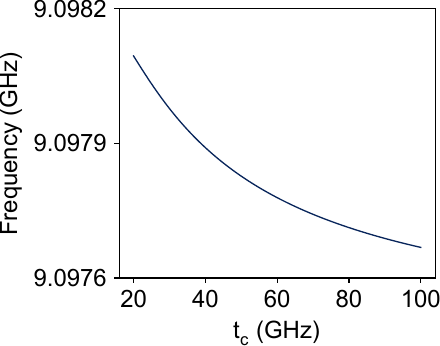}
\caption{\textbf{Simulated dispersion of the 01 transition frequency with respect to the interdot tunnel coupling. } It is computed with the parameters used to reproduce the time domain measurement. }
\label{tunnel_simu}
\end{figure}

\bibliographystyle{apsrev4-2}

\makeatother

\end{document}